\documentclass[twocolumn,aps,prd,showpacs,superscriptaddress,preprintnumbers,amsmath,amssymb]{revtex4}

\pdfoutput=1
\newcommand{\gsim}{\mathrel{\rlap{\raisebox{.3ex}{$>$}}
    \raisebox{-.6ex}{$\sim$}}}
\newcommand{\lsim}{\mathrel{\rlap{\raisebox{.3ex}{$<$}}
    \raisebox{-.6ex}{$\sim$}}}

\usepackage{url}
\usepackage{graphicx}
\usepackage[english]{babel}
\usepackage[thinspace,mediumqspace]{SIunits}
\usepackage{multirow}

\graphicspath{{./figs/}}

\begin{document}

\title{\bf Obtaining supernova directional information using the neutrino matter oscillation pattern}
\author{Kate Scholberg}\affiliation{Department of Physics, Duke University, Durham, NC 27708 USA}
\author{Armin Burgmeier}\affiliation{Universit\"at Karlsruhe, 76128 Karlsruhe, Germany}
\author{Roger Wendell}\affiliation{Department of Physics, Duke University, Durham, NC 27708 USA}

\date{\today}

\begin{abstract}

  A nearby core collapse supernova will produce a burst of neutrinos
  in several detectors worldwide.  With reasonably high probability,
  the Earth will shadow the neutrino flux in one or more detectors.
  In such a case, for allowed oscillation parameter scenarios, the
  observed neutrino energy spectrum will bear the signature of
  oscillations in Earth matter.  Because the frequency of the
  oscillations in energy depends on the pathlength traveled by the
  neutrinos in the Earth, an observed spectrum contains also
  information about the direction to the supernova.  
  We explore here the possibility of constraining the supernova
  location using matter oscillation patterns observed in a 
  detector.  Good energy
 resolution (typical of scintillator detectors), well known oscillation parameters, and optimistically large (but 
 conceivable) statistics are required. 
 Pointing by this method can be significantly improved
  using multiple detectors located around the globe.  Although it is
  not competitive with neutrino-electron elastic scattering-based pointing
  with water Cherenkov detectors,
  the technique could still be useful.

\end{abstract}

\maketitle

\section{Introduction}\label{intro}

The core collapse of a massive star leads to emission of a short, intense burst of
neutrinos of all flavors.   The time scale is tens of seconds and the neutrino energies are in
the range of a few tens of MeV.  Several detectors worldwide, both current and planned for the near future, are  sensitive to a core collapse burst within the Milky Way or slightly beyond~\cite{Scholberg:2007nu}.

The first electromagnetic radiation is not expected to emerge from the star for hours, or perhaps even a few days.  Therefore any directional information that can be extracted from the neutrino signal will be advantageous to astronomers who can use such information to initiate a search for the visible supernova.  We note that not every core collapse may produce a bright supernova: some supernovae may be obscured, and some core collapses may produce no
supernova at all, in which case directional information will aid the search for a remnant.

The possibility of using the neutrinos themselves to point back to the
supernova has been explored in the literature~\cite{Beacom:1998fj,
  Tomas:2003xn}.  Triangulation  based on relative timing of neutrino
burst signals was also considered in~\cite{Beacom:1998fj}; however
available statistics, as well as considerable practical difficulties
in prompt sharing of information, makes time triangulation more difficult.
Leaving
aside the possibility of a TeV neutrino
signal~\cite{Tomas:2003xn} (which would likely be delayed), the most
promising way of using the neutrinos to point to a supernova is via
neutrino-electron elastic scattering: neutrinos interacting with atomic
electrons scatter their targets within a cone of about 25$^\circ$ with
respect to the supernova direction.  The quality of pointing goes as
$\sim N^{-1/2}$, where $N$ is the number of elastic scattering events.
In water and scintillator detectors,  neutrino-electron elastic scattering
represents only a few percent of the total signal, which is dominated
by inverse beta decay $\bar{\nu}_e + p \rightarrow n + e^+$, 
for which anisotropy is
weak~\cite{Vogel:1999zy}.  Furthermore the directional information in
the elastic
scattering signal is available only for water Cherenkov detectors, for
which direction information is preserved via the Cherenkov cone of the
scattered electrons.
Taking into account the near-isotropic background of non-elastic scattering
events~\cite{Tomas:2003xn}, a Super-K-like detector~\cite{Ikeda:2007sa} (22.5~kton
fiducial volume) will have 68\% (90\%) C.L.  pointing of about 6$^\circ$  (8$^\circ$)
for
a 10~kpc supernova; this could improve to $< 1^\circ$ for next-generation Mton-scale
water detectors.
Long string water detectors~\cite{Halzen:1995ex} do not
reconstruct supernova neutrinos event-by-event and so cannot use this channel
for pointing.
Scintillation light is nearly isotropic and so scintillation detectors
have very poor directional capability, although there is potentially
information in the relative positions of the inverse beta decay positron and neutron
vertices~\footnote{This technique was studied for gadolinium-loaded
  scintillator in references~\cite{Apollonio:1999jg,Hochmuth:2007gv}; however
  gadolinium loading for future large scintillator detectors does not
  seem to be planned.}, and some novel scintillator directional
techniques are under development~\cite{nutech}.

We consider here a new possibility: detectors with sufficiently good
energy resolution will be able to obtain directional information by
observing the effects of neutrino oscillation on the energy spectrum
of the observed neutrinos, assuming that oscillation parameters are
such that matter oscillations are present.  Although not
competitive with elastic scattering, some directional
information can be obtained even in a single detector (unlike for
time triangulation).  Combinations of detectors at different locations
around the globe may yield fairly high quality information.

\section{Determining the Direction with Earth Matter Effects}

Supernova neutrinos traversing the Earth's matter before reaching a
detector will experience matter-induced oscillations, depending on the
values of the MNS matrix parameters ~\cite{Dighe:1999bi,Lunardini:2001pb,
Takahashi:2001dc,
Lunardini:2003eh,Dighe:2003jg,Dighe:2003vm}.
Whether or not there will be an Earth matter effect depends on
currently-unknown mixing parameters, $\theta_{13}$ and the mass
hierarchy: matter oscillation will occur for both $\nu_e$ and
$\bar{\nu}_e$ for values of $\sin^2\theta_{13}\lsim 10^{-5}$, for
normal but not inverted hierarchy; if $\theta_{13}$ is relatively
large, $\sin^2\gsim 10^{-3}$, then matter oscillation occurs for
$\bar{\nu}_e$ but not $\nu_e$ for either
hierarchy~\cite{Dasgupta:2008my,Dighe:2008dq}.  The frequency of the
oscillation in $L/E$, where $E$ is the neutrino energy and $L$ is the
neutrino pathlength in Earth matter, depends on now fairly
well-known mixing parameters.  Therefore, the oscillation
pattern in neutrino energy $E$ measured at a single detector contains
information about the pathlength $L$ traveled through the Earth
matter.  If the pathlength $L$ is known, one knows that supernova is
located somewhere on a ring on the sky corresponding to this
pathlength.  If another pathlength is measured at a different location
on the globe, the location can be further constrained to the intersection of the
allowed regions.

A Fourier transform of the inverse-energy
distribution~\cite{Dighe:2003jg} of the observed neutrinos will yield
a peak if oscillations are present.  References~\cite{Dighe:2003jg}
and \cite{Dighe:2003vm} explore the conditions under which peaks are
observable with a view to obtaining information about the oscillation
parameters.  The authors assume that the direction of the supernova,
and hence the pathlength through the Earth, is known.  Here we turn the argument
around: we assume that oscillation parameters are such that the matter
effects do occur and can be identified, and that enough is known about
MNS parameters to extract information about $L$ and hence about
supernova direction from the data.  A similar idea to determine
possible georeactor location from the oscillated spectrum was explored
in reference~\cite{Dye:2009jp}.  We note that by the time a nearby
supernova happens, the hierarchy and whether $\theta_{13}$ is large or
small may in fact be known from long-baseline and reactor experiments.
With reasonably high probability~\cite{Mirizzi:2006xx}, the Earth will
shadow the supernova in at least one detector.  We note that
\textit{lack} of observation of a matter peak in the inverse-energy
transform (assuming there should be one) gives some direction information
as well: if no peak is present, one can infer that
the supernova is overhead at a
given location.  If the hierarchy and value of $\theta_{13}$ are
already known with sufficient precision 
at the time of the supernova, we should know in advance
whether or not a peak in the $k$ distribution should appear;
otherwise, its appearance for at least one detector location may
answer the question.

\section{Evaluation of the Concept in Idealized Scenarios}

To evaluate the general feasibility of this concept we make several
simplifying assumptions.  We consider only inverse beta decay in large
water Cherenkov and liquid scintillator detectors 
(we ignore the presence of other interactions, which should be a small
correction; some of them can be tagged)
\footnote{Large liquid argon detectors will also have supernova
  neutrino sensitivity~\cite{Bueno:2003ei}.  Such detectors are
  primarily sensitive to $\nu_e$ rather than $\bar{\nu}_e$; they have
  good energy resolution and in principle could employ the matter
  oscillation pointing technique for the case when mixing parameters
  favor $\nu_e$ oscillation in the Earth.  However liquid argon time
  projection chambers also have excellent intrinsic pointing
  capability, and the angular resolution for neutrino-electron elastic
  scattering will certainly be superior. Therefore we will not
  consider liquid argon further here.}. 
We will first consider a detector with perfect energy resolution, and
then consider resolutions more typical of real water Cherenkov and scintillator
detectors.

We borrow some of the assumptions and notation of reference~\cite{Dighe:2003jg}.
We assume a neutrino
interaction cross-section proportional to $E^2$, perfect detection
efficiency above threshold and no background.
We assume a ``pinched'' neutrino spectrum of the form:
$F_0 = \frac{\phi_0}{E_0}  \frac{(1 + \alpha)^{1+\alpha}}{\Gamma\left(1+\alpha\right)}  \left(\frac{E}{E_0}\right)^\alpha  \mathrm e^{-\left(\alpha+1\right)\frac{E}{E_0}}$, where $E_0$ is the average neutrino energy.  We choose
parameters 
$\alpha = 3$, average energies for the flavors
$E_{\bar{\nu}_e} = \unit{15}{\mega\electronvolt}$ and
$E_{\bar{\nu}_x} = \unit{18}{\mega\electronvolt}$, and
$\frac{\phi_{\bar{\nu}_e}}{\phi_{\bar{\nu}_x}} = 0{.}8$.  These parameters correspond to
the ``Garching'' model~\cite{Raffelt:2003en}.
We ignore for this study ``spectral splits'' (\textit{e.g.} \cite{Dasgupta:2009mg}) or
other features which will introduce additional Fourier 
components.    We assume that there are no non-standard neutrino
interactions or other exotic effects that modify the spectra.

The oscillation probabilities have been computed by numerical solution
of the matter oscillation equations~\cite{roger} using these vacuum
parameters and the full PREM Earth density model~\cite{PREM}. Between
neighboring radial points in the model the matter density is taken to
be constant such that the three-neutrino transition amplitude may be
computed following the methods outlined in~\cite{Barger80}. The final
amplitude is the product of all amplitudes across the matter slices
along the neutrino's trajectory. The initial flux of neutrinos is
taken to arrive at the Earth as pure mass states such that the
detection probability is taken according to the probability of a neutrino being
$\bar{\nu}_e$ flavor
when it reaches the detector.
The oscillation parameters were chosen to be
$\sin^2{2\theta_{12}} = 0{.}87$,
$\sin^2{2\theta_{13}} = 0$,
$\sin^2{2\theta_{23}} = 1{.}0$,
$\Delta m_{12}^2 = 7{.}6 \times \unit{10^{-5}}{\electronvolt^2}$, and
$\Delta m_{23}^2 = 2{.}4 \times \unit{10^{-3}}{\electronvolt^2}$.

\subsection{Perfect Energy Resolution}

\begin{figure}[!htbp]
\begin{minipage}{3in}
\includegraphics[width=3.2in]{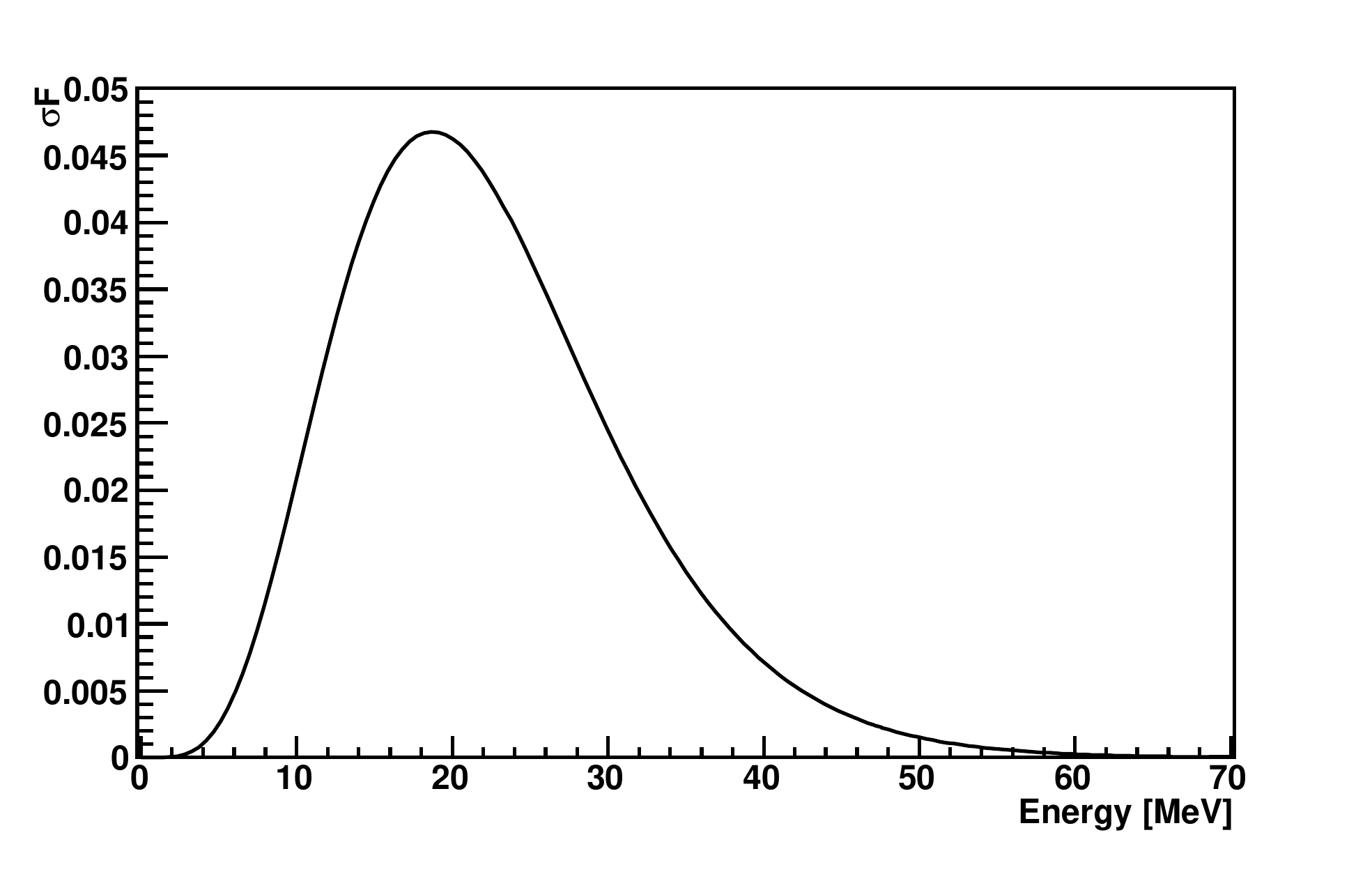}
\end{minipage}
\begin{minipage}{3in}
\includegraphics[width=3.2in]{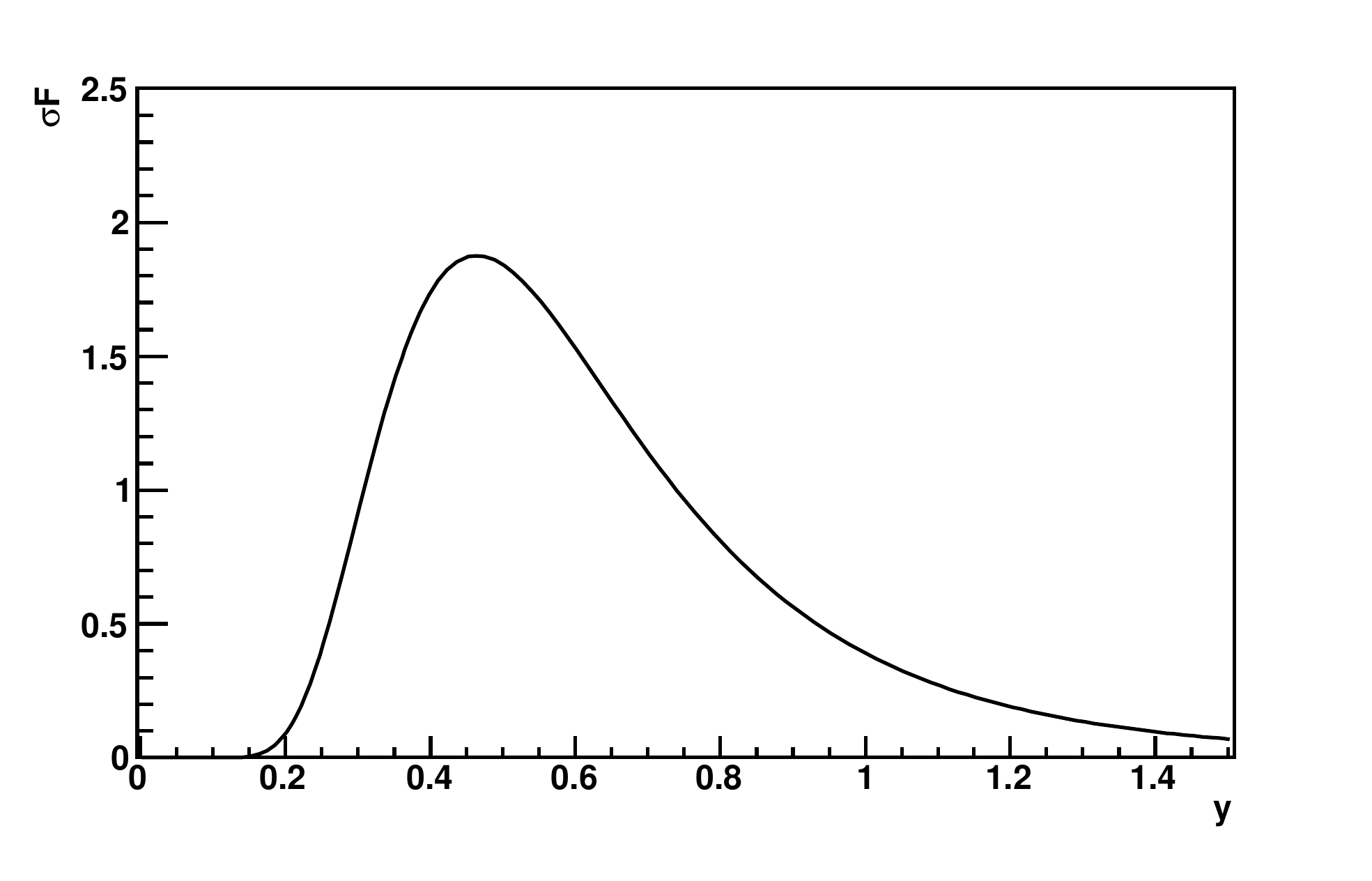}
\end{minipage}
\caption{Top: assumed neutrino event spectrum without oscillations.  Bottom: inverse-energy distribution.}
\label{fig:nooscil}
\end{figure}

\begin{figure}[!htbp]
\begin{minipage}{3in}
\includegraphics[width=3.2in]{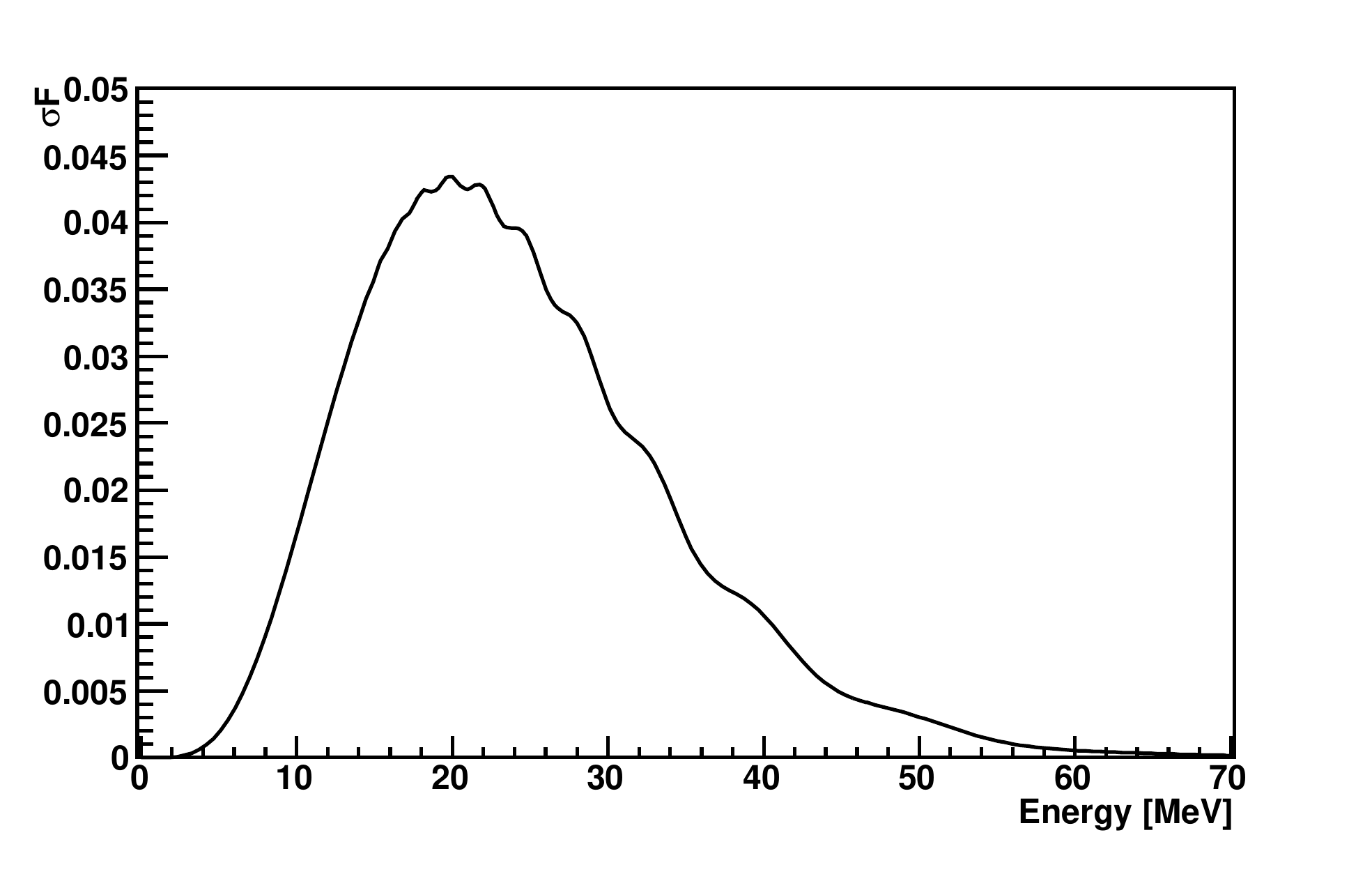}
\end{minipage}
\begin{minipage}{3in}
\includegraphics[width=3.2in]{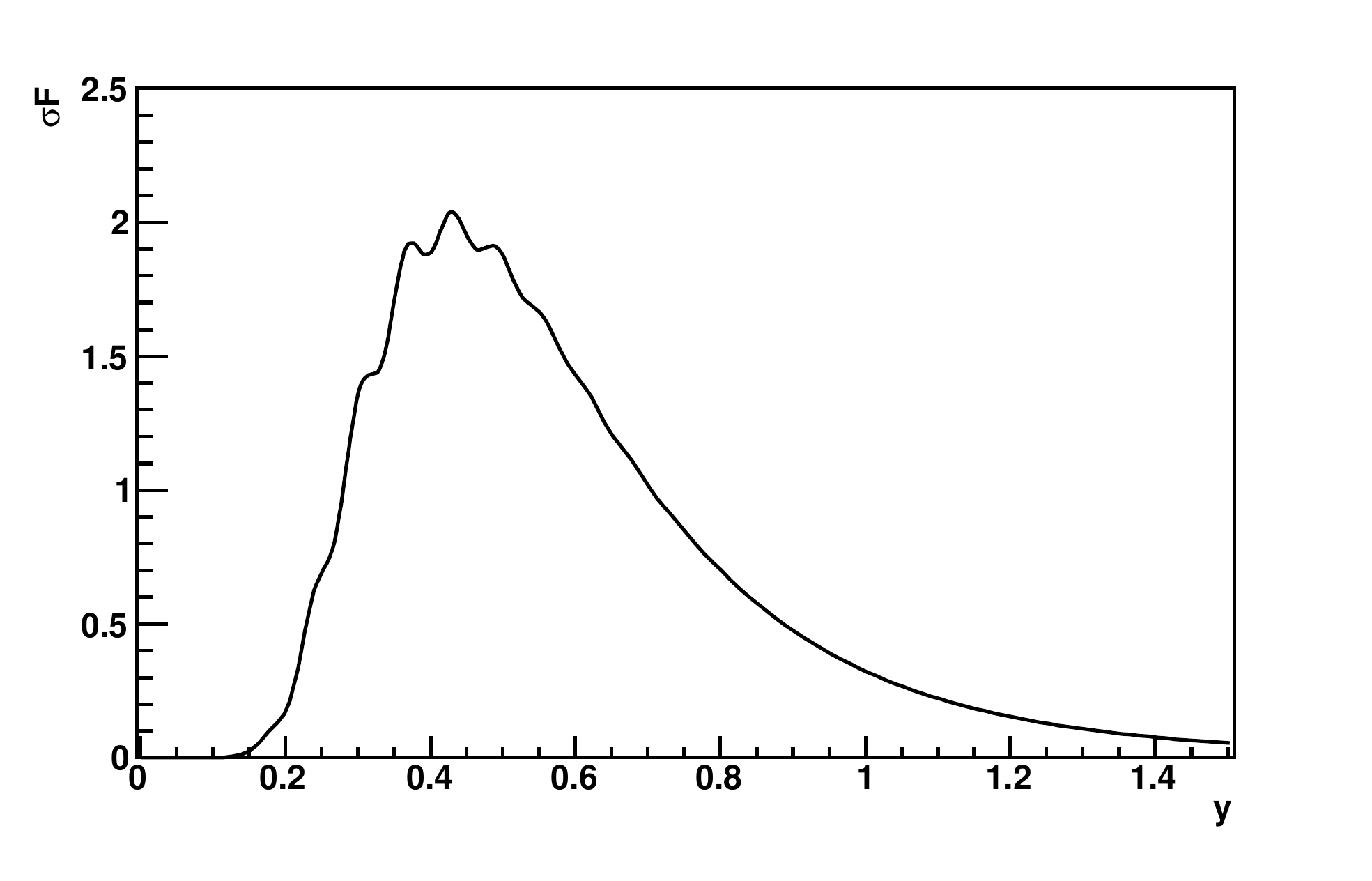}
\end{minipage}
\caption{Top: assumed neutrino event spectrum with matter oscillations for $L=6,000$~km.  Bottom: inverse-energy distribution with matter oscillations.}
\label{fig:oscil}

\end{figure}

The spectrum of inverse beta decay events, integrated over time, 
is shown in Fig.~\ref{fig:nooscil}.   Fig.~\ref{fig:nooscil} 
shows on the bottom the ``inverse-energy'' spectrum, where the inverse-energy
parameter $y$ is defined as $y = \frac{\unit{12{.}5}{\mega\electronvolt}}{E}$.
 Fig.~\ref{fig:oscil} shows the Earth matter modulation of the spectrum, for
$L=6,000$~km.  Shown on the bottom is the modulation in inverse-energy, for which the
peaks are evenly spaced.

\begin{figure}[!htbp]
\begin{minipage}{3in}
\includegraphics[width=3.2in]{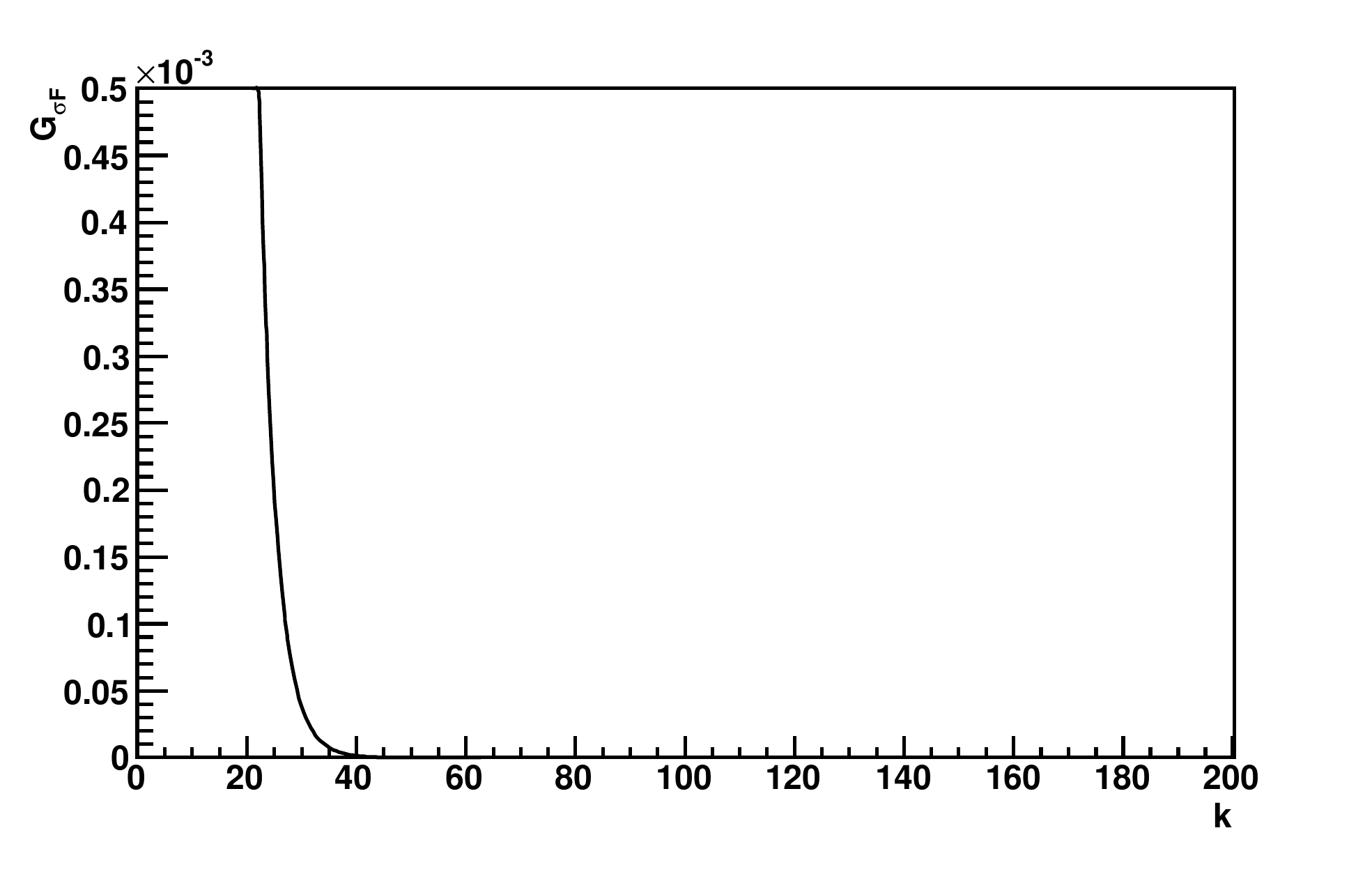}
\end{minipage}
\begin{minipage}{3in}
\includegraphics[width=3.2in]{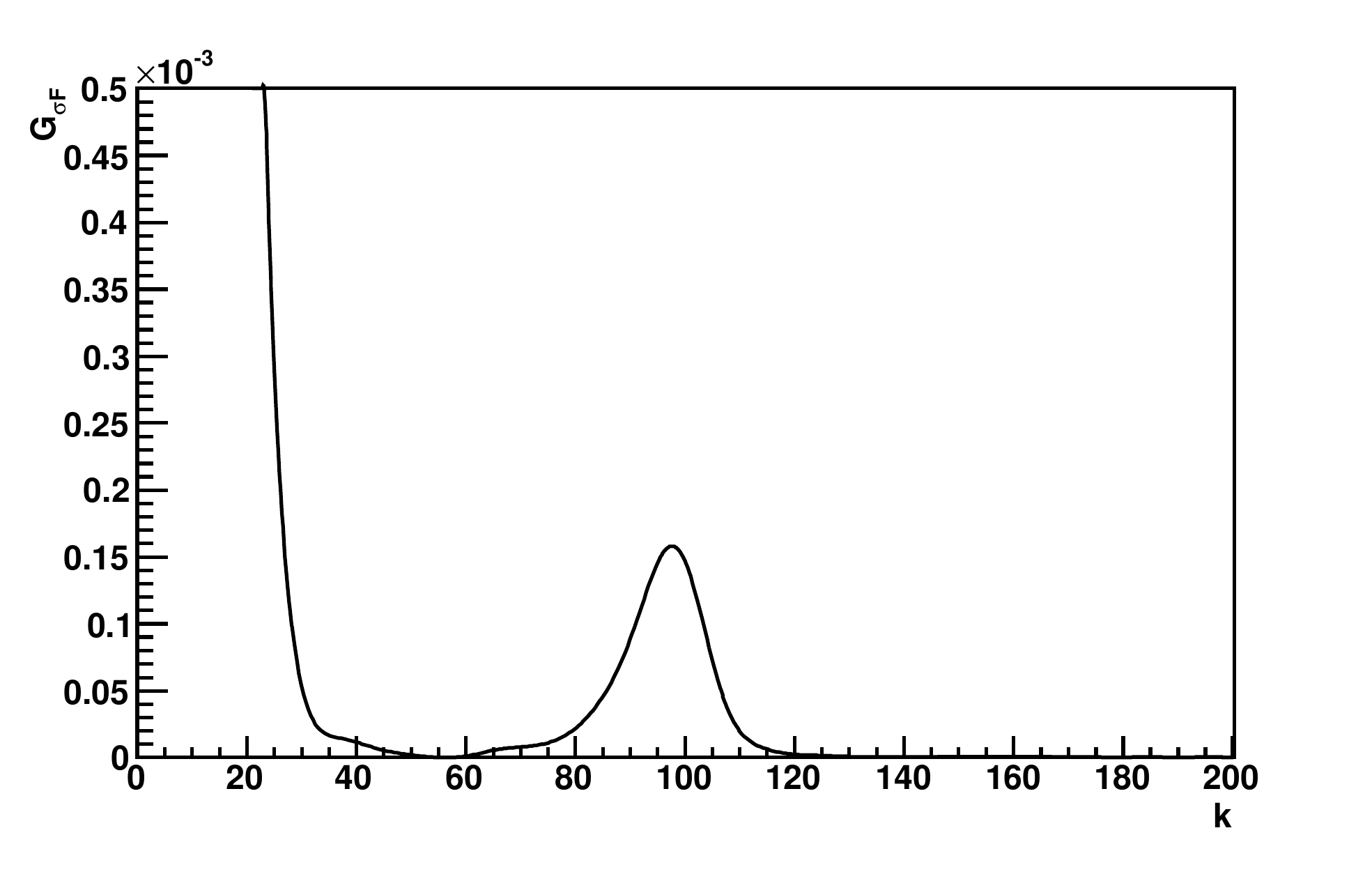}
\end{minipage}
\caption{Inverse-energy power spectrum without (top) and with (bottom) matter oscillations. }
\label{fig:fourier}
\end{figure}

\begin{figure}[!htbp]
\begin{centering}
\includegraphics[width=3.2in]{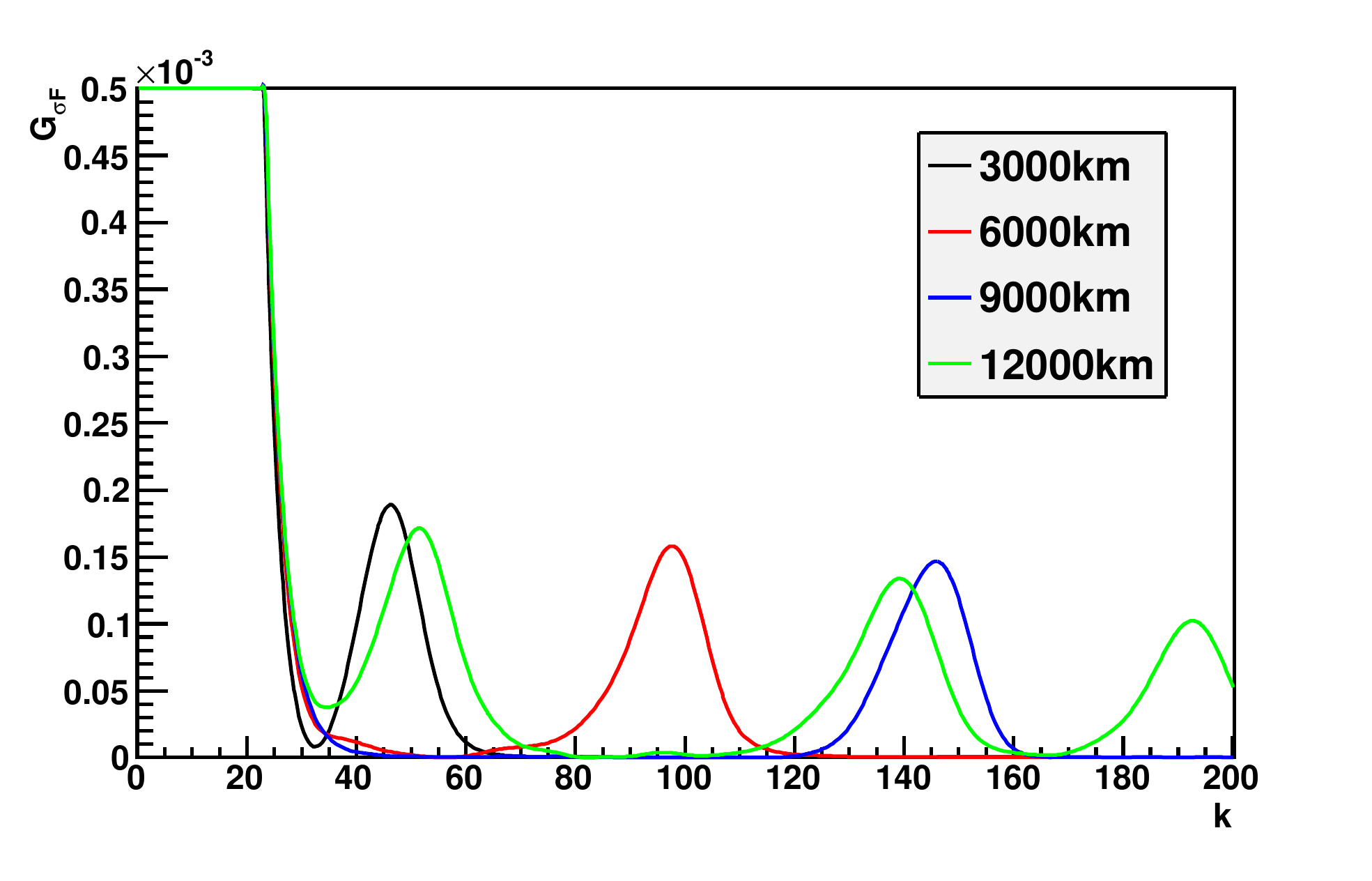}
\caption{Examples of inverse-energy power spectra for several pathlengths.}
\label{fig:fourier_peaks}
\end{centering}
\end{figure}

The Fourier transform of the detected
inverse-energy spectrum is $g(k)=\int_{-\infty}^{\infty} f(y) e^{iky} dy$.
The power spectrum $G_{\sigma F}(k)=|g(k)|^2$ assuming perfect energy resolution is shown in Fig.~\ref{fig:fourier},  for no matter oscillation on the top and for matter oscillation on the bottom, assuming pathlength $L=6,000$~km.
The power spectra are generated from the normalized inverse-energy
distributions for which $\int_0^\infty \sigma F(y) \mathrm dy =
1$.  Thus the power spectra are normalized so that $G_{\sigma F}(0) =
1$.
Fig.~\ref{fig:fourier_peaks} shows the power spectra for several values of $L$, illustrating how the peak moves
to higher $k$ values as the pathlength increases.  For pathlengths such that the
neutrinos traverse the Earth core ($L>10,700$~km), additional peaks are present in the spectrum~\cite{Dighe:2003vm}.
There is no observable peak for $L$ less than about 2500~km, for which the neutrinos are no
longer traversing much high-density matter.

\begin{figure}[!htbp]
\begin{centering}
\includegraphics[width=3.2in]{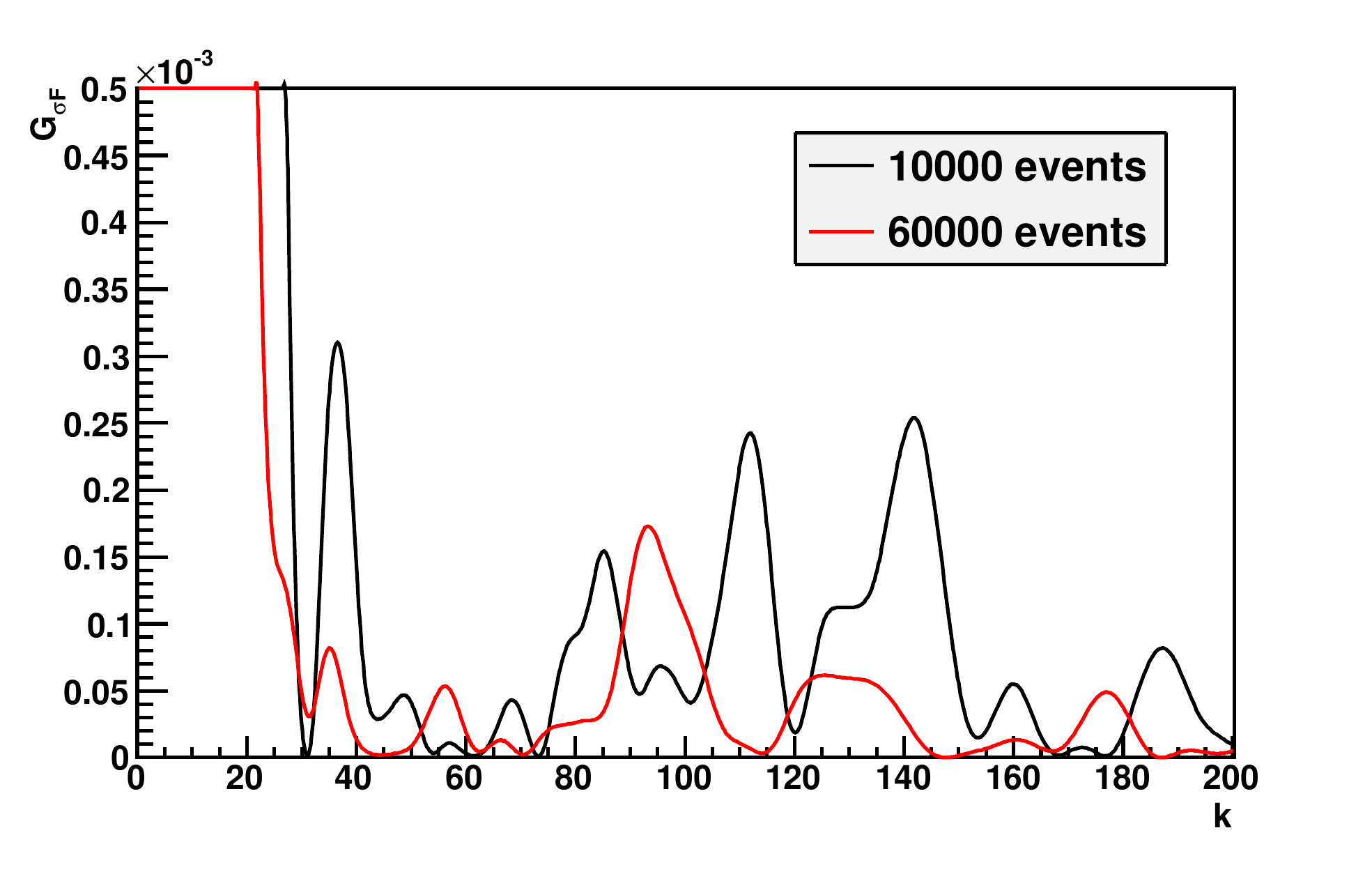}
\caption{Examples of inverse-energy power spectra for perfect energy resolution but finite statistics. Both lines show $L=6000$~km.}
\label{fig:fourier_peaks_finite_stats}
\end{centering}
\end{figure}

Fig.~\ref{fig:fourier_peaks_finite_stats} shows now the effect of
finite statistics, for a simulated supernova with 10,000 events and one with
60,000 events.  The finite statistics result in a background for the
main peak(s) in the power spectrum.  For most of the following, we consider a
rather optimistically large (but not unthinkable) 60,000 event signal,
which would correspond to a supernova at a distance of about 5~kpc 
observed with a 50~kton detector.

\subsubsection{Method for Determining Directional Information}

If one measures $k_{\rm peak}$, the position of the largest 
peak in the power spectrum, for a supernova signal, one can in
principle determine the pathlength traveled by the neutrino in the
Earth.  We use a simple Neyman construction method~\cite{Amsler:2008zzb}
to estimate the quality of directional information.

\begin{figure}[!htbp]
\begin{centering}
\includegraphics[width=3.2in]{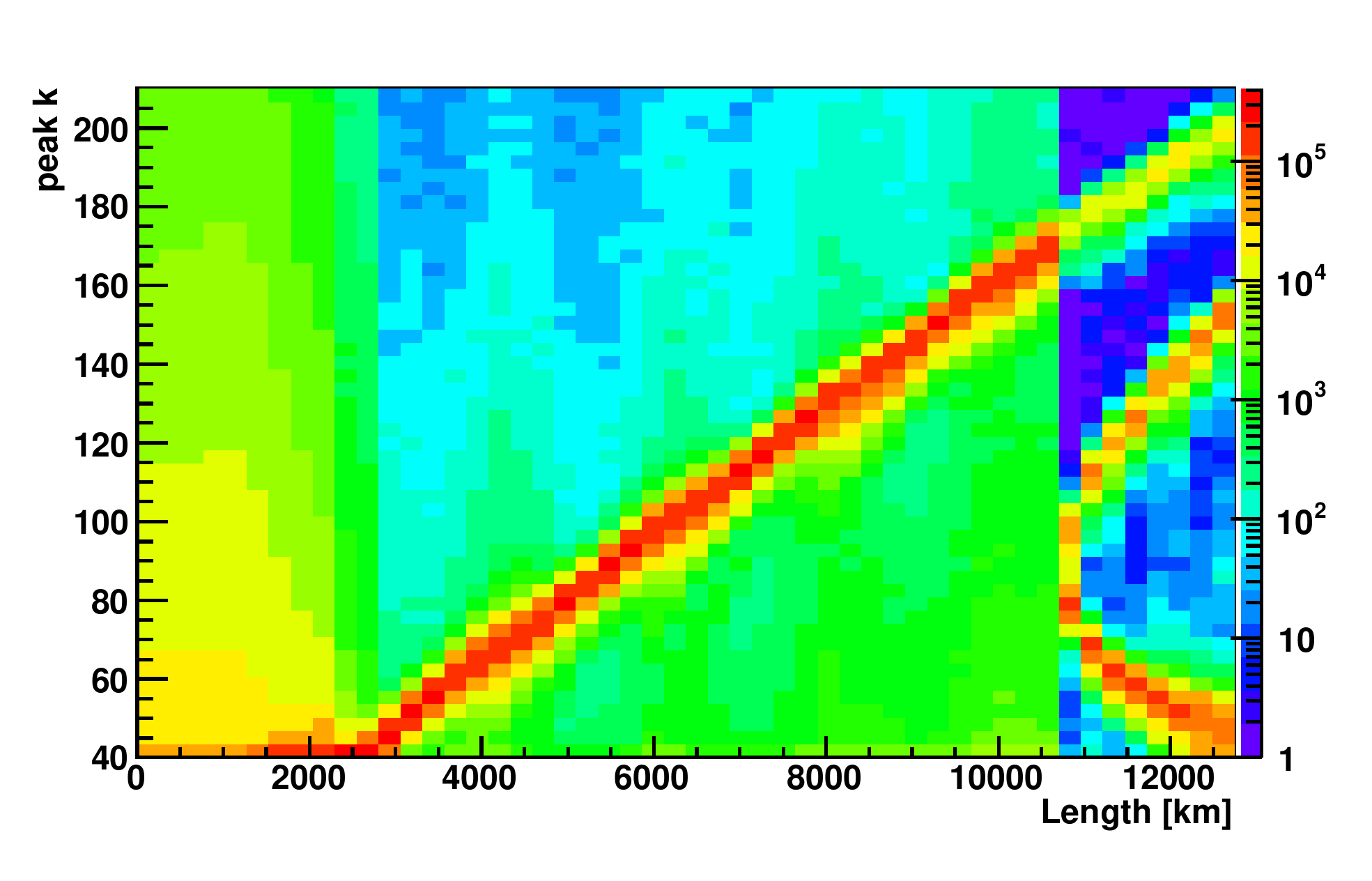}
\caption{ Distribution of the position of the maximum peak in $k$ as a function of matter-traversed pathlength $L$, assuming perfect energy resolution. There are 500,000 simulated supernovae per $L$, each with 60,000 events.}
\label{fig:peak_distances_perfect}
\end{centering}
\end{figure}

We first find the position of the largest peak in $k$ as a function
of pathlength $L$, assuming perfect energy resolution but finite
statistics.  To find the peak in the power spectrum, we first set a
lower threshold of $k=40$ and an upper threshold of $k=210$. Below
that threshold, the peak merges with the low $k$ peak (corresponding
to the unoscillated spectrum) and can no longer be identified.  Peaks
beyond $k=210$ would correspond to distances greater than the diameter
of the Earth.  For each $k$ within that range we then evaluate the
integral from $k-\Delta k/2$ to $k+\Delta k/2$ which corresponds to
the area under the peak. We take the $k$ for which this value is
highest as the peak in the spectrum. We chose $\Delta k=4$. Even
though Fig.~\ref{fig:fourier_peaks} suggests that peaks can be wider
than that, we found more fluctuation in the peak's position for higher
$\Delta k$ when taking finite energy resolution into account,
especially for small distances ($L < 4000$~km).
Fig.~\ref{fig:peak_distances_perfect} shows that the value of $k_{\rm
  peak}$ is clearly correlated with pathlength $L$; for distances less
than about 2000~km, for which the neutrinos do not undergo matter
oscillations, it represents mainly random noise. 
The multiple peak structure for neutrinos passing through the
core is clearly visible for $L>10,700~$km.  We note that the height
of the largest peak also contains information about $L$, as do the
secondary peak positions, if such exist.

\begin{figure}[!htbp]
\begin{centering}
\includegraphics[width=3.2in]{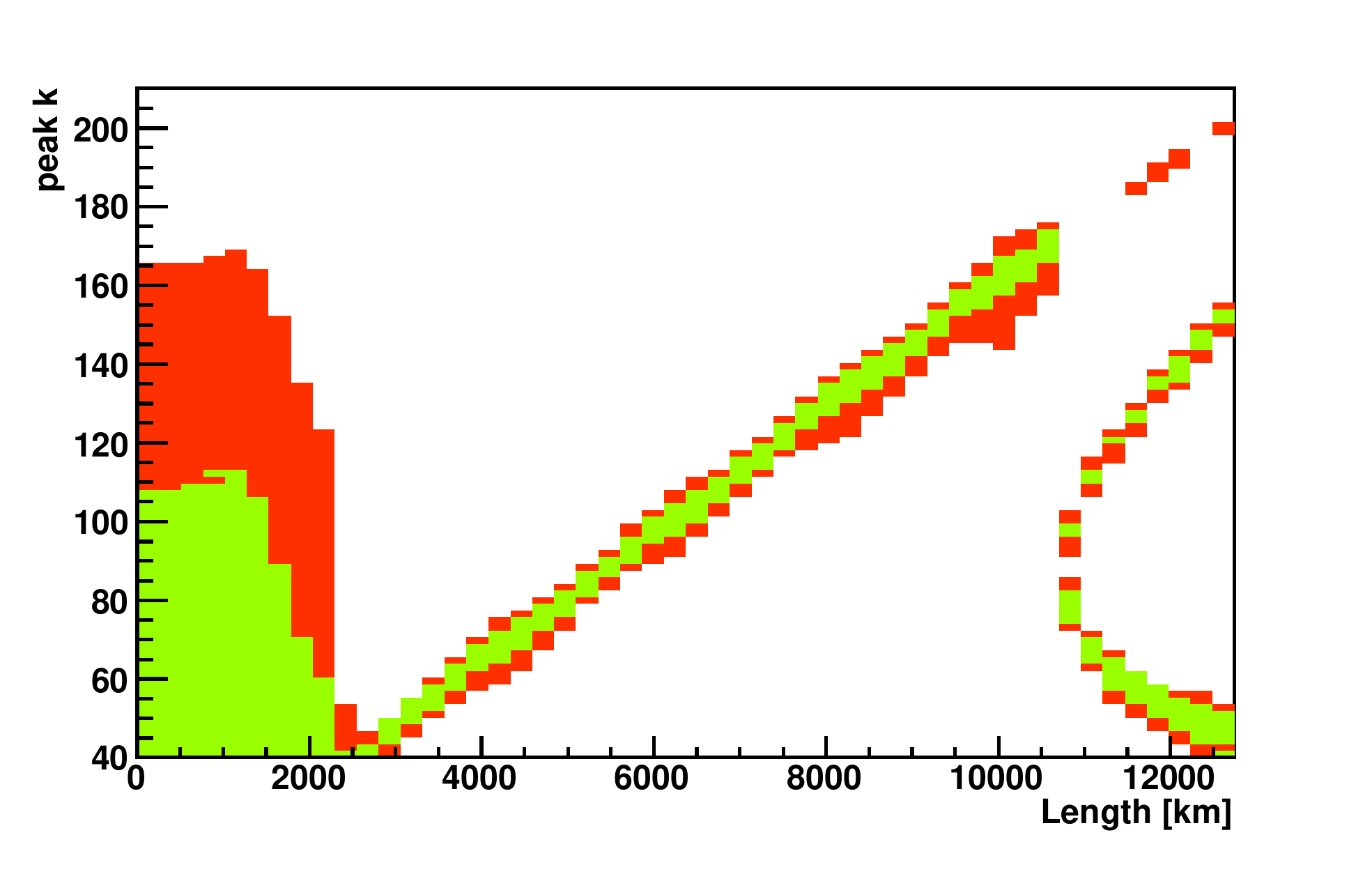}
\caption{Neyman construction for $k_{\rm peak}$ and $L$: for a given measured $k_{\rm peak}$ one reads off a range of allowed $L$ values. The green area shows the 68\% confidence region and the red area the 90\% one.}
\label{fig:neyman_perfect}
\end{centering}
\end{figure}

Given a particular measurement of $k_{\rm peak}$, one can then
determine a range of distances $L$ allowed, making use of the Neyman
construction shown in Fig.~\ref{fig:neyman_perfect}. To ensure contiguous
regions in $k$ we drop regions that contribute less than 3\% to the final
Neyman construction and increase existing regions instead so that the total
covered area is 68\% or 90\%.
The range in $L$ values can then be mapped to an allowed region on the sky.
We have checked explicitly that the statistical coverage is as expected.

\begin{figure}[!htbp]
\begin{minipage}{3in}
\includegraphics[width=3.2in]{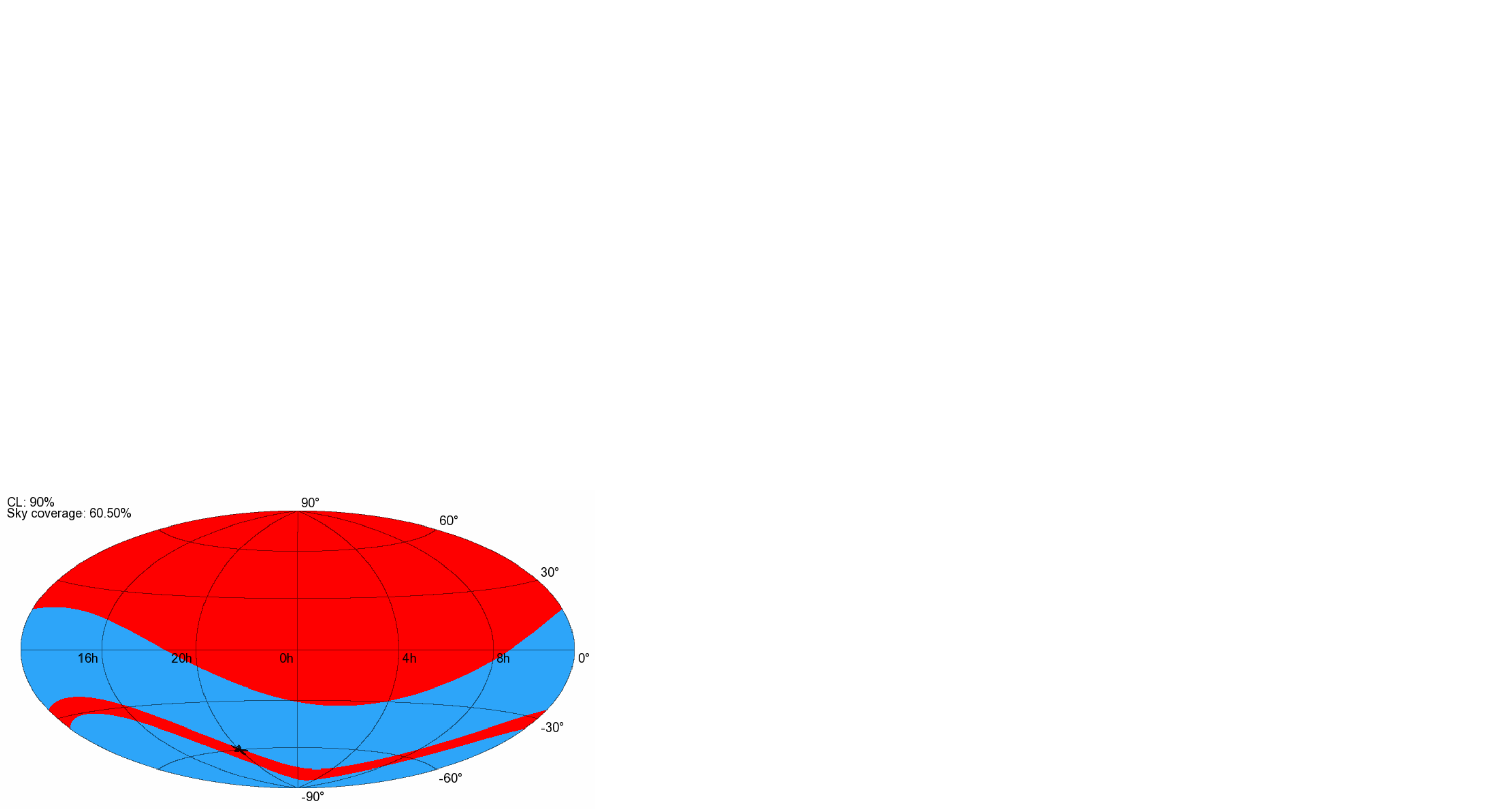}
\end{minipage}
\begin{minipage}{3in}
\includegraphics[width=3.2in]{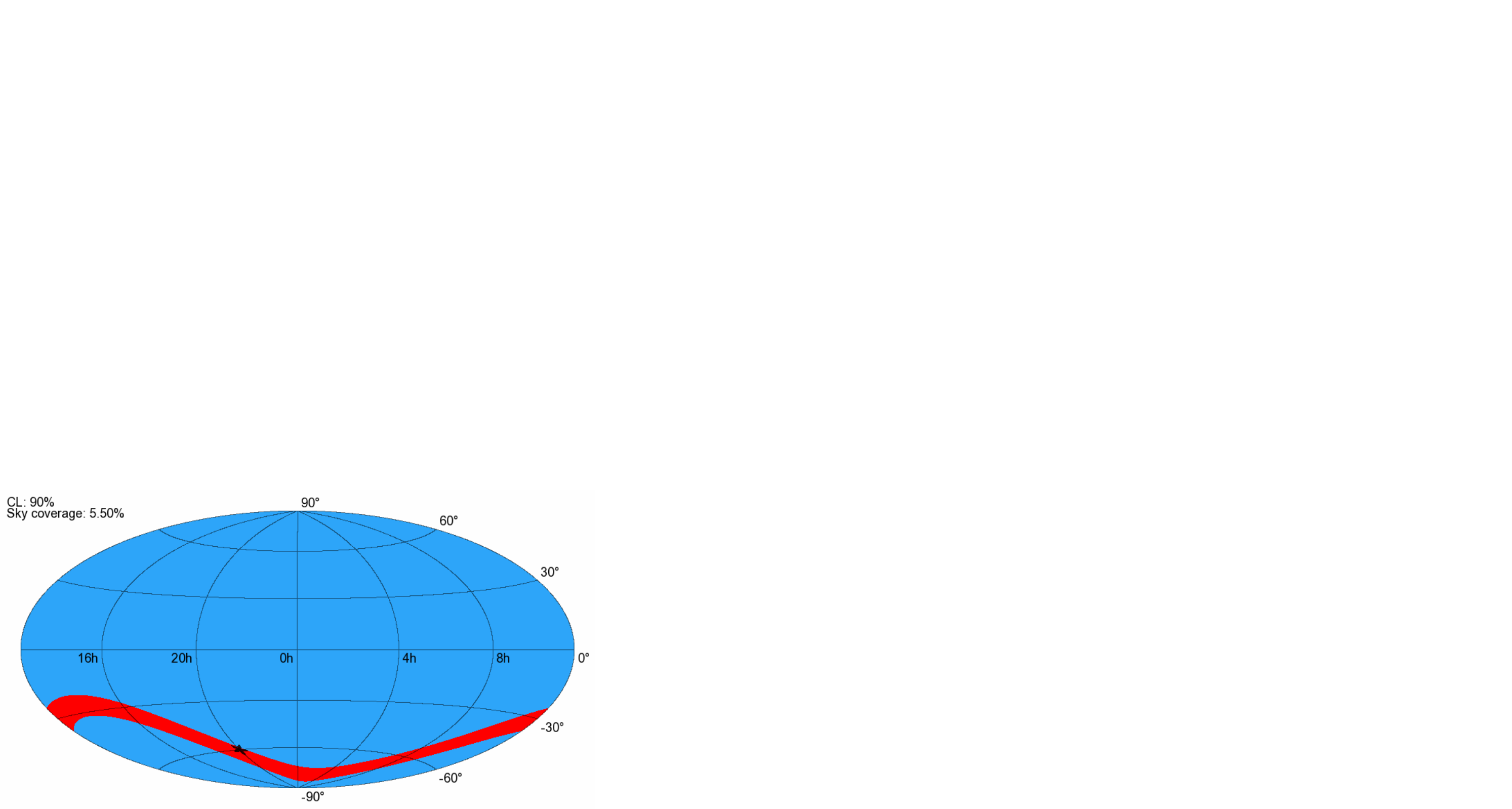}
\end{minipage}

\caption{Example Hammer projection skymaps in equatorial coordinates, showing 90\% C. L. allowed regions on the sky for a supernova at the position indicated by a star.
A 60,000 neutrino event signal measured in Finland with perfect energy resolution is assumed.  The top plot shows the allowed region without taking into account peak height; the bottom plot takes into account peak height information. }
\label{fig:skymap_perfect}
\end{figure}

Fig.~\ref{fig:skymap_perfect} shows an example Hammer projection sky
map in equatorial coordinates showing 90\% C.L. allowed regions for
an assumed true supernova direction (indicated by a star) of R.A.$=20^h$
and decl.$=-60^\circ$ (occurring at 0:00 GMST), for assumed perfect energy resolution and
statistics of 60,000 events.

\begin{figure}[!htbp]
\begin{centering}
\includegraphics[width=3.2in]{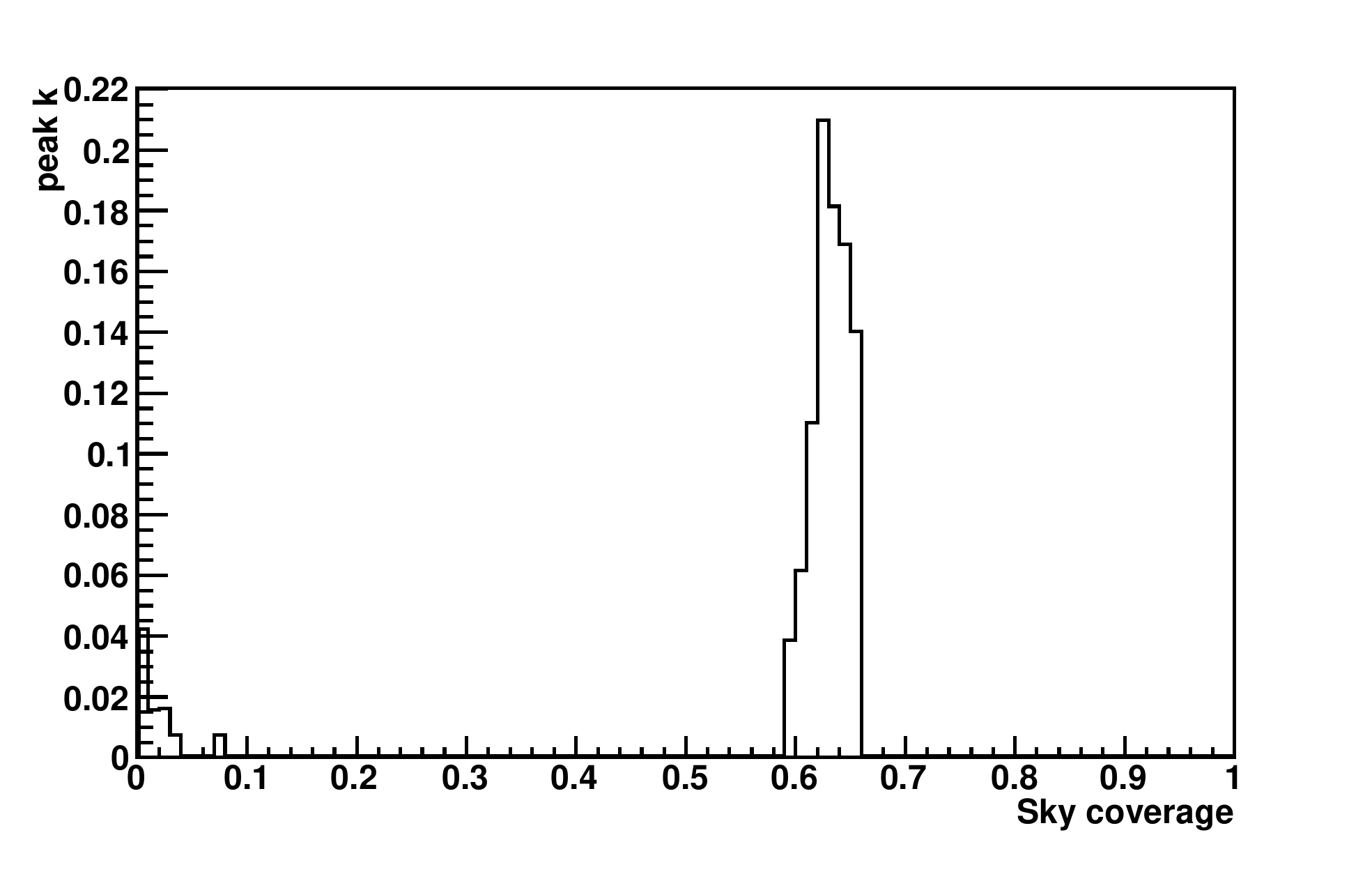}
\caption{Histogram of fractional sky coverages for the 90\% C.L. region, assuming perfect energy resolution in a single detector, using $k_{\rm peak}$ information only, for
the example configuration of Fig.~\ref{fig:skymap_perfect}.}
\label{fig:sky_coverage_all_perfect}
\end{centering}
\end{figure}

\begin{figure}[!htbp]
\begin{centering}
\includegraphics[width=3.2in]{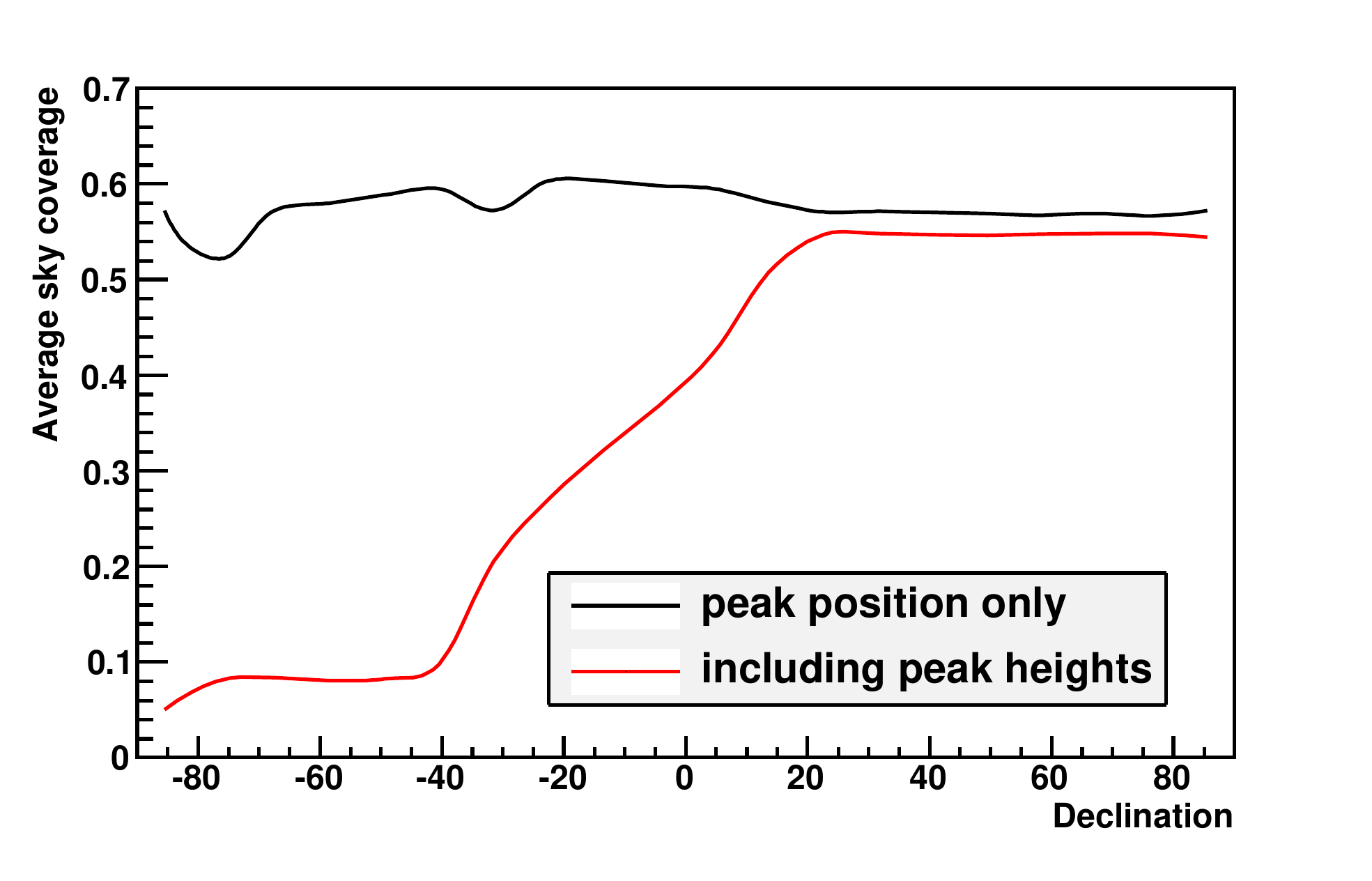}
\caption{Sky coverage averaged over right ascension as a function of declination, for a single detector.  The black line just takes into account the peak position, whereas the red line includes also the height of the peak.  In total, 83,500 supernovae, evenly distributed over declination, have been simulated for both lines.}
\label{fig:sky_coverage_vs_declination_perfect}
\end{centering}
\end{figure}

Fig.~\ref{fig:sky_coverage_all_perfect} shows the distribution of fractional sky coverage for perfect energy resolution.  The distribution is bimodal, because the
$L<2500$~km possibility (corresponding to large fractional sky coverage)
is often not excluded at 90\% in the Neyman construction.  
Fig.~\ref{fig:sky_coverage_vs_declination_perfect} shows the average
sky coverage vs. declination of the supernova, averaged over 24h of right ascension,
for a detector located in Finland ($63{.}66^\circ$ N, $26{.}04^\circ$ E).

\begin{figure}[!htbp]
\begin{centering}
\includegraphics[width=3.2in]{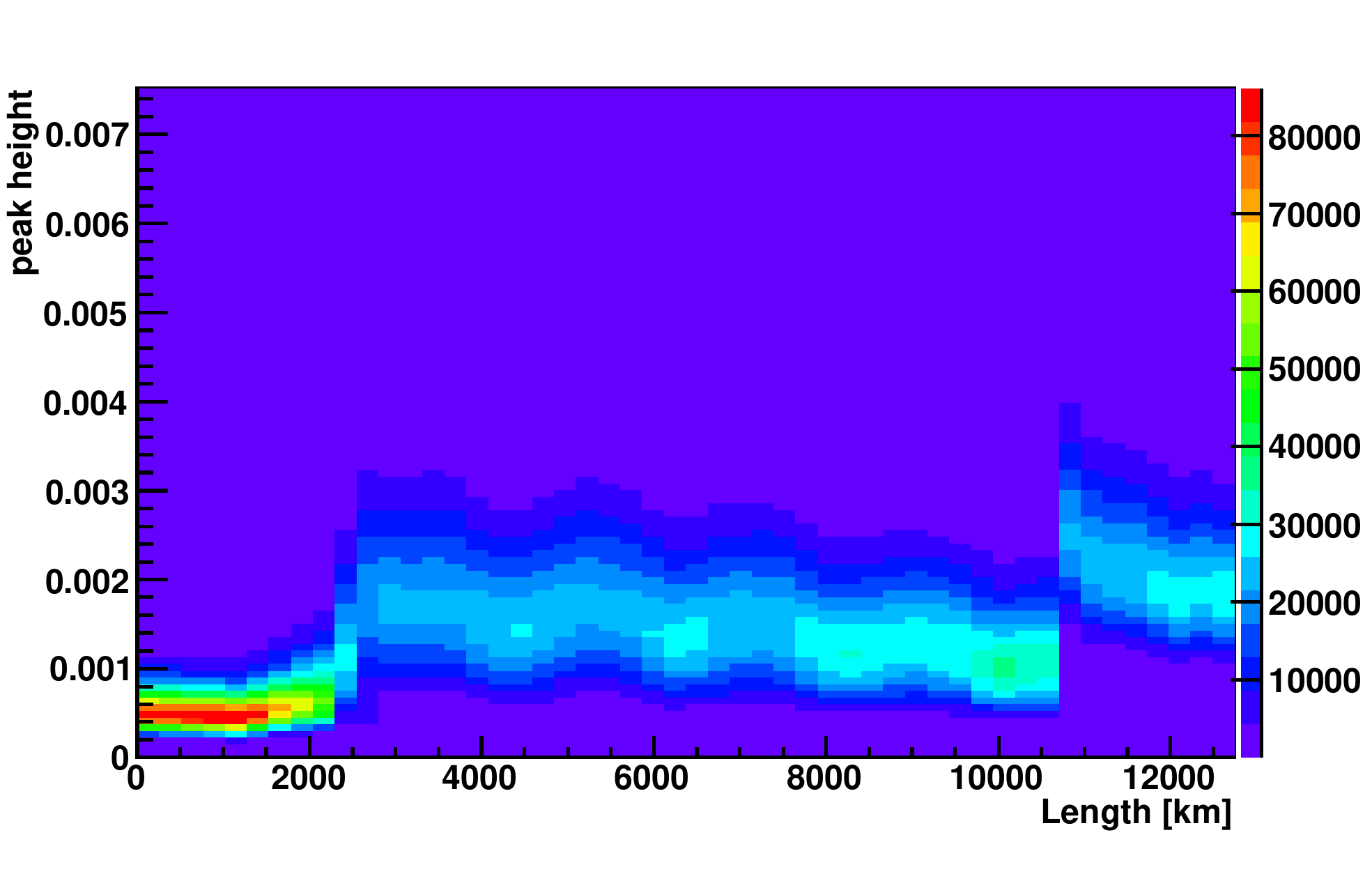}
\caption{Distribution of the heights of the maximum peak in $k$ as a function of matter-traversed pathlength $L$, assuming perfect energy resolution. There are 500,000 simulated supernovae per $L$, each with 60,000 events.}
\label{fig:peak_heights_perfect}
\end{centering}
\end{figure}

If we incorporate also information about the height of the largest peak $h$ 
into a Neyman construction, for long pathlengths we can remove the
possibility of a short-pathlength overhead supernova, and improve the
pointing quality significantly.  Fig.~\ref{fig:peak_heights_perfect} shows
the correlation between peak heights and $L$. Figs.~\ref{fig:skymap_perfect} (bottom) and
\ref{fig:sky_coverage_vs_declination_perfect} show the effect of incorporating this
information.
Subsequent plots will assume use of both Fourier 
peak position and height information.

\subsubsection{Combining Detectors}

Clearly, having several detectors around the globe observing the
neutrino burst will improve the measurement.  
If each of the
detectors could select a single $L$, an observation with two
detectors will produce two allowed regions where the rings on the sky
overlap, and a third observation will narrow it down to one spot.
However because more than one $L$ region may be allowed for a
given detector, the combination can include multiple regions.

For the multiple detector case, we make the
Neyman construction for 100,000 randomly chosen
$(k_1;h_1,k_2;h_2,\ldots)$-tuples only (with 100 bins in $k_{\rm peak}$ and
height for each detector) in order to compute it in a reasonable
amount of time. In this case the smoothing procedure described above
is not applied.

\begin{figure}[!htbp]
\begin{minipage}{3in}
\includegraphics[width=3.2in]{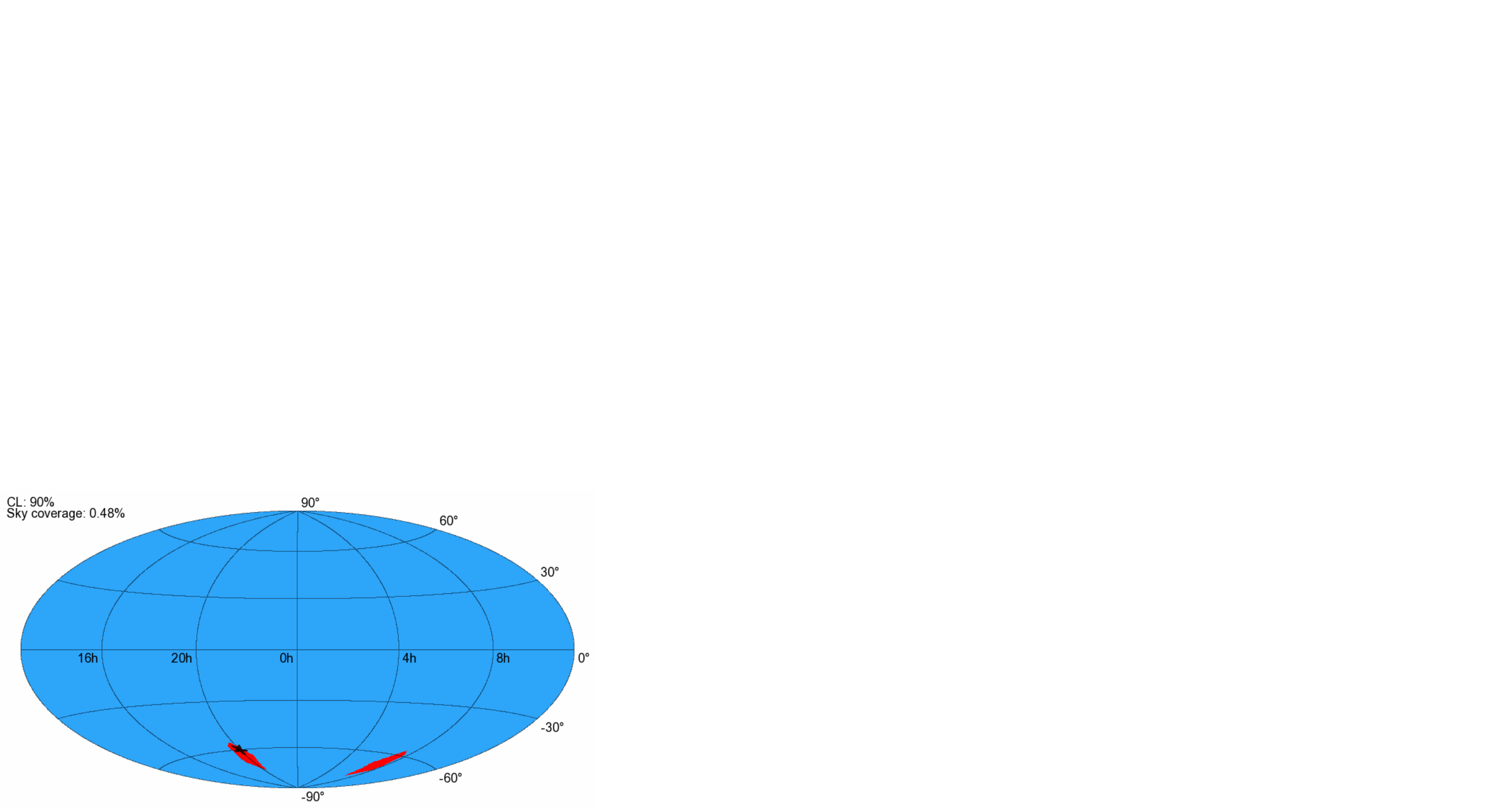}
\end{minipage}
\begin{minipage}{3in}
\includegraphics[width=3.2in]{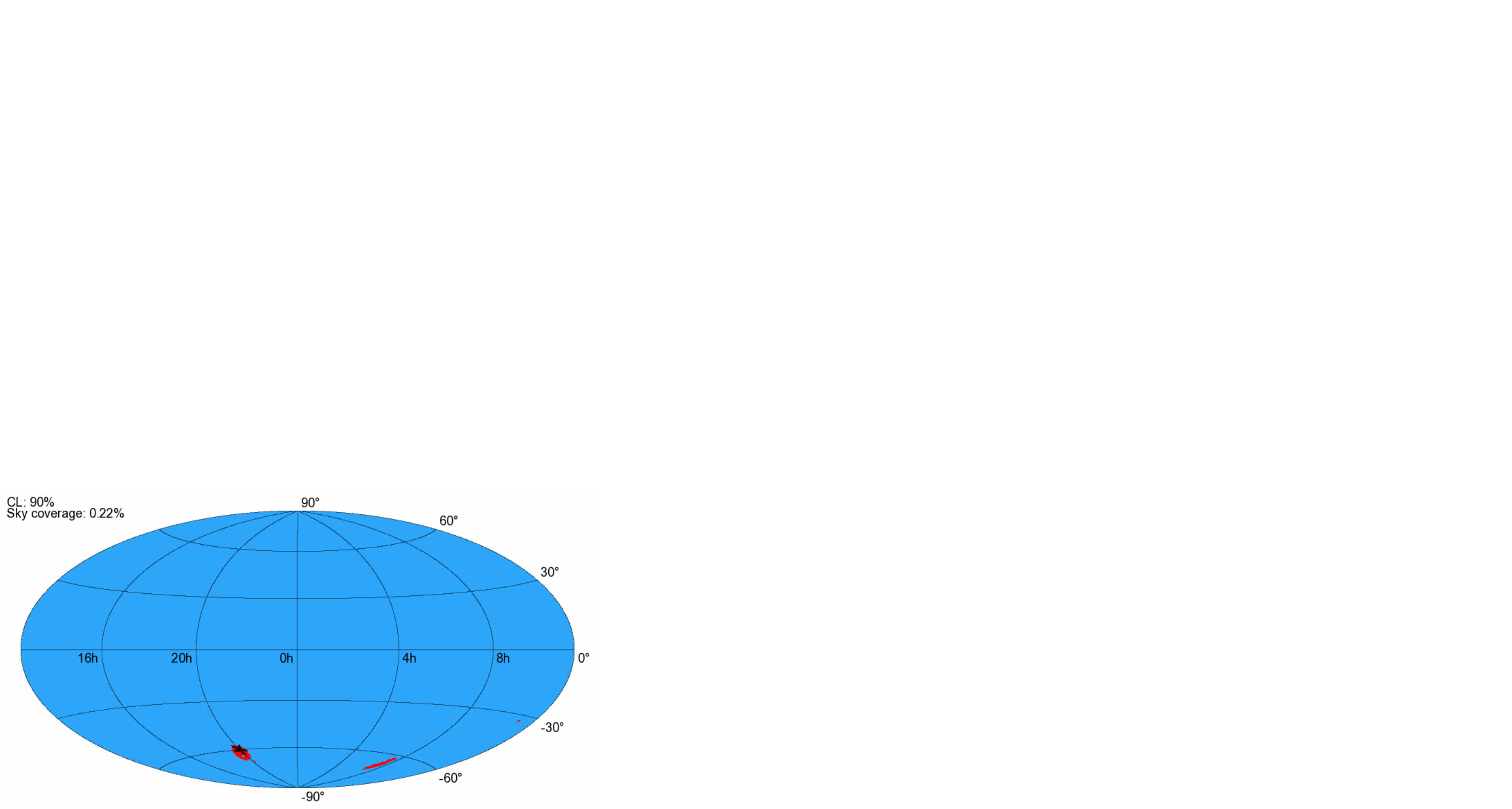}
\end{minipage}

\caption{Combined skymaps for detectors with perfect resolution. Top: Two detectors with 60,000 events each.   Bottom: Three detectors with 60,000 events each.}
\label{fig:combined_skymap}
\end{figure}

\begin{figure}[!htbp]
\begin{centering}
\includegraphics[width=3.2in]{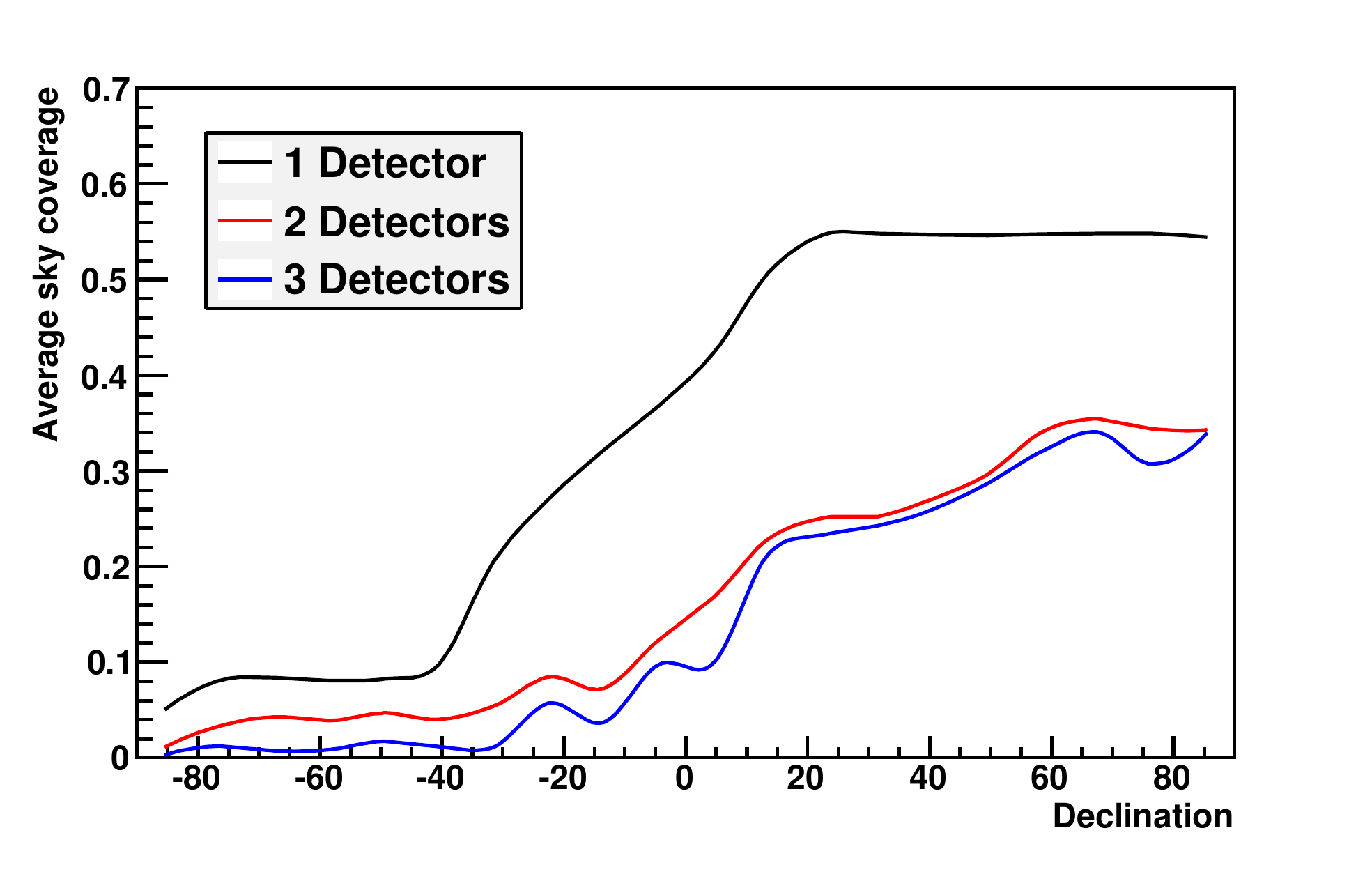}
\caption{Sky coverage averaged over right ascension as a function of declination, for one, two and three detectors. For one detector 83,500 supernovae have been simulated, for two detectors the number is 2,630 and for three detectors it is 760.}
\label{fig:multiple_sky_coverage_vs_declination_perfect}
\end{centering}
\end{figure}

Fig.~\ref{fig:combined_skymap} shows example scenarios involving two and three
detectors and Fig.~\ref{fig:multiple_sky_coverage_vs_declination_perfect} summarizes
average quality as a function of declination.
Clearly in this idealized situation, 
combined information is quite good, and the more
detectors spread around the globe, the better.

\subsection{More Realistic Detectors}

\begin{figure}[!htbp]
\begin{centering}
\includegraphics[width=3.2in]{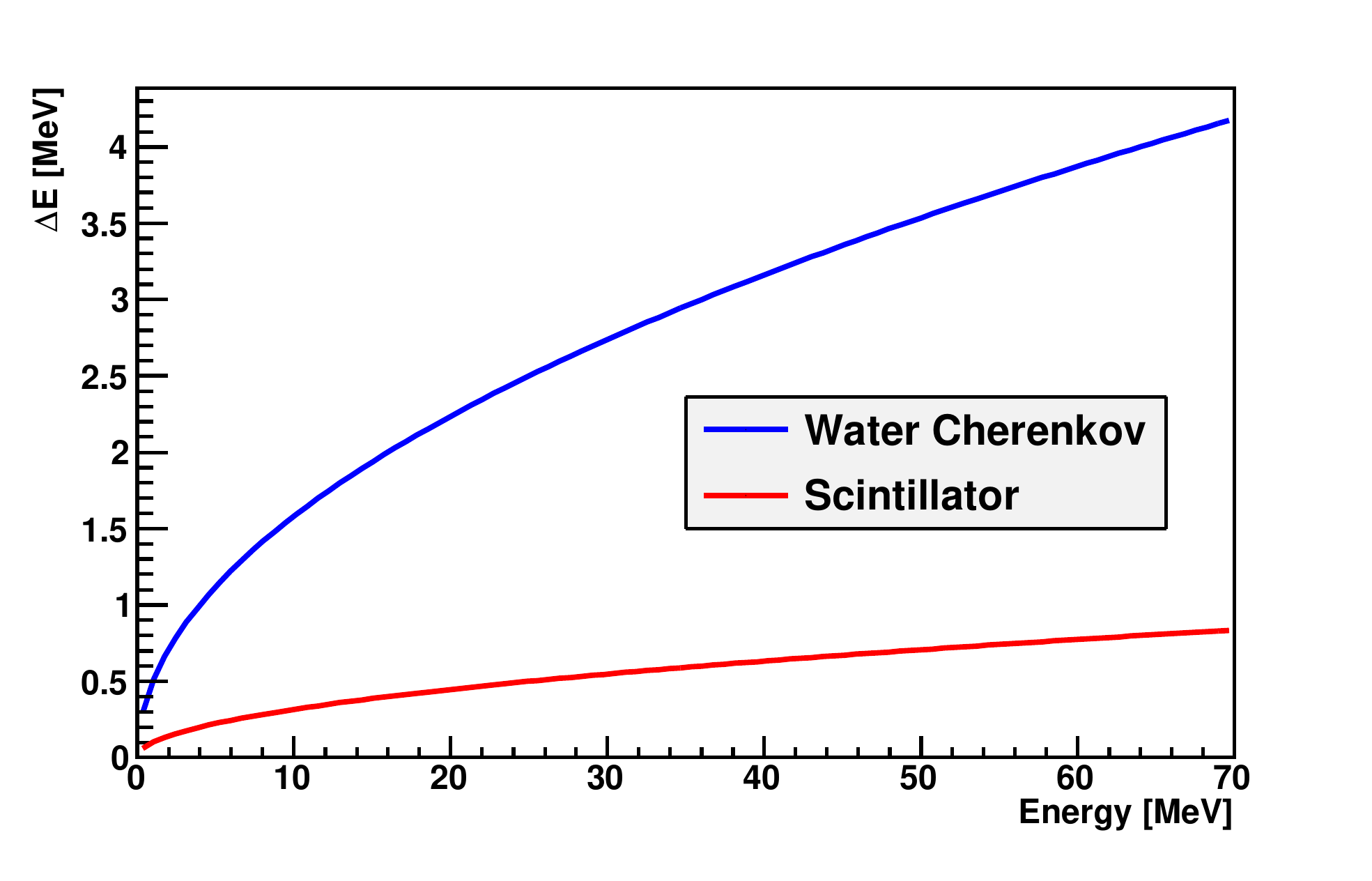}
\caption{Energy resolution functions used for water and scintillator. $\Delta E$ is the standard deviation $\sigma$ of a Gaussian.}
\label{fig:energy_resolution}
\end{centering}
\end{figure}

Next we will assume a slightly more realistic situation.  Imperfect
energy resolution will tend to smear out the oscillation pattern and
degrade the detectability of the peak in $k$.  We estimate the effect
of energy resolution by selecting events from the spectrum and
smearing their energies according to a Gaussian of the prescribed
width.  The energy resolution functions used,  
the same as in reference~\cite{Dighe:2003vm}, are shown in
Fig.~\ref{fig:energy_resolution}; one is characteristic of
scintillator and one of water Cherenkov detectors.  For water
Cherenkov we assume a threshold of 5~MeV and for scintillator we
assume a threshold of 1~MeV.

\subsubsection{Water Cherenkov Detectors}

\begin{figure}[!htbp]
\begin{centering}
\includegraphics[width=3.2in]{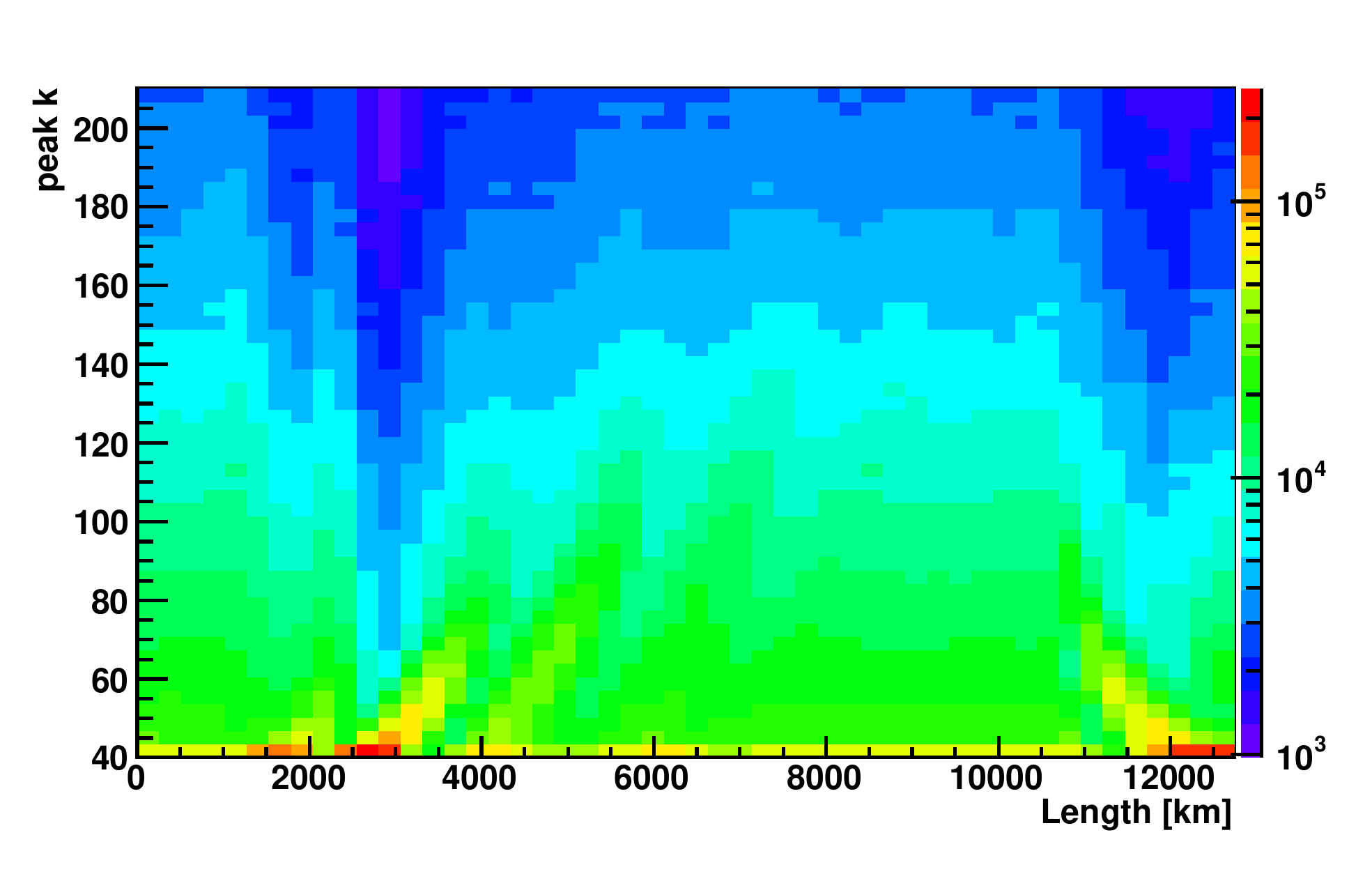}
\caption{ Distribution of the position of the maximum peak in $k$ as a function of matter-traversed pathlength $L$, assuming water Cherenkov energy resolution. There are 500,000 simulated supernovae per $L$, each with 60,000 events.}
\label{fig:peak_distances_water}
\end{centering}
\end{figure}

\begin{figure}[!htbp]
\begin{centering}
\includegraphics[width=3.2in]{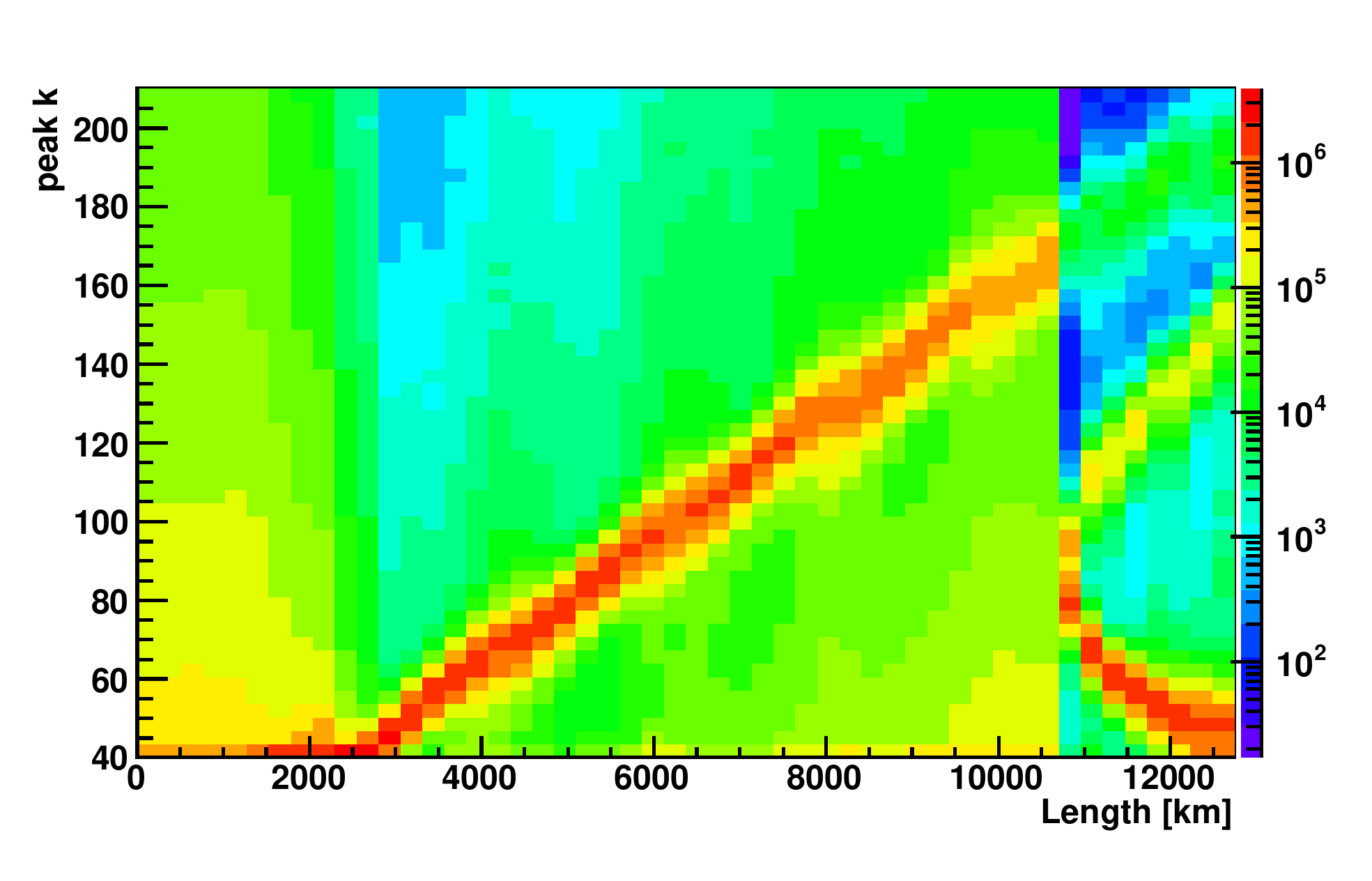}
\caption{ Distribution of the position of the maximum peak in $k$ as a function of matter-traversed pathlength $L$, assuming scintillator energy resolution. There are 5,000,000 simulated supernovae per $L$, each with 60,000 events.}
\label{fig:peak_distances_scint}
\end{centering}
\end{figure}

Fig.~\ref{fig:peak_distances_water} 
shows the distribution of 
$k_{\rm peak}$ and $L$  for simulated supernovae in a 
detector with water-Cherenkov-like energy resolution.  
Fig.~\ref{fig:peak_distances_scint} shows the same for scintillator.

Clearly the water Cherenkov resolution smears the power spectrum
information enough to preclude its use for this purpose; furthermore,
far superior direction information will come from elastic scattering
in a water Cherenkov detector.  Therefore we will focus subsequent
attention on scintillator detectors, which have significantly better
energy resolution and weak intrinsic direct pointing capabilities.

\subsubsection{Scintillator Detectors}

Existing and near-future scintillator detectors with supernova
neutrino detection capabilities are KamLAND~\cite{Eguchi:2002dm},
LVD~\cite{Aglietta:1992dy,Agafonova:2006fz}, Borexino~\cite{Cadonati:2000kq} and
SNO+~\cite{Kraus:2006qp}; these are however probably too small
to acquire the large statistics required for this technique.
Future scintillator detectors of the tens of kton scale for which this
technique could be feasible are LENA~\cite{MarrodanUndagoitia:2008zz},
to be sited in Finland, and the ocean-based HanoHano~\cite{Learned:2008zj}.

\begin{figure}[!htbp]
\begin{minipage}{3in}
\includegraphics[width=3.2in]{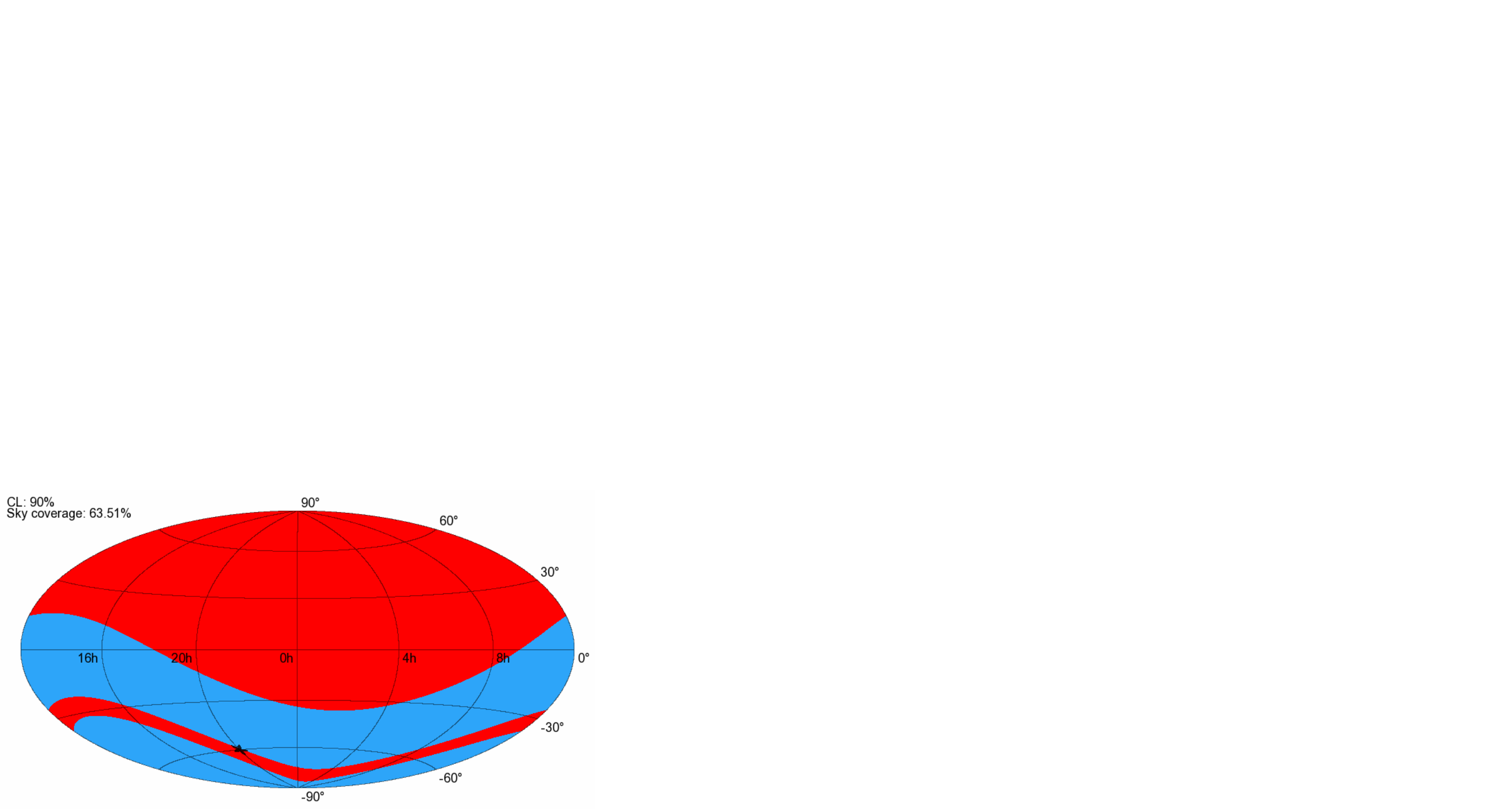}
\end{minipage}
\begin{minipage}{3in}
\end{minipage}

\caption{Example scintillator skymaps, for a single detector located in 
Finland, assuming a 60,000 event signal (top). }
\label{fig:scintillator_skymap}
\end{figure}

\begin{figure}[!htbp]
\begin{centering}
\includegraphics[width=3.2in]{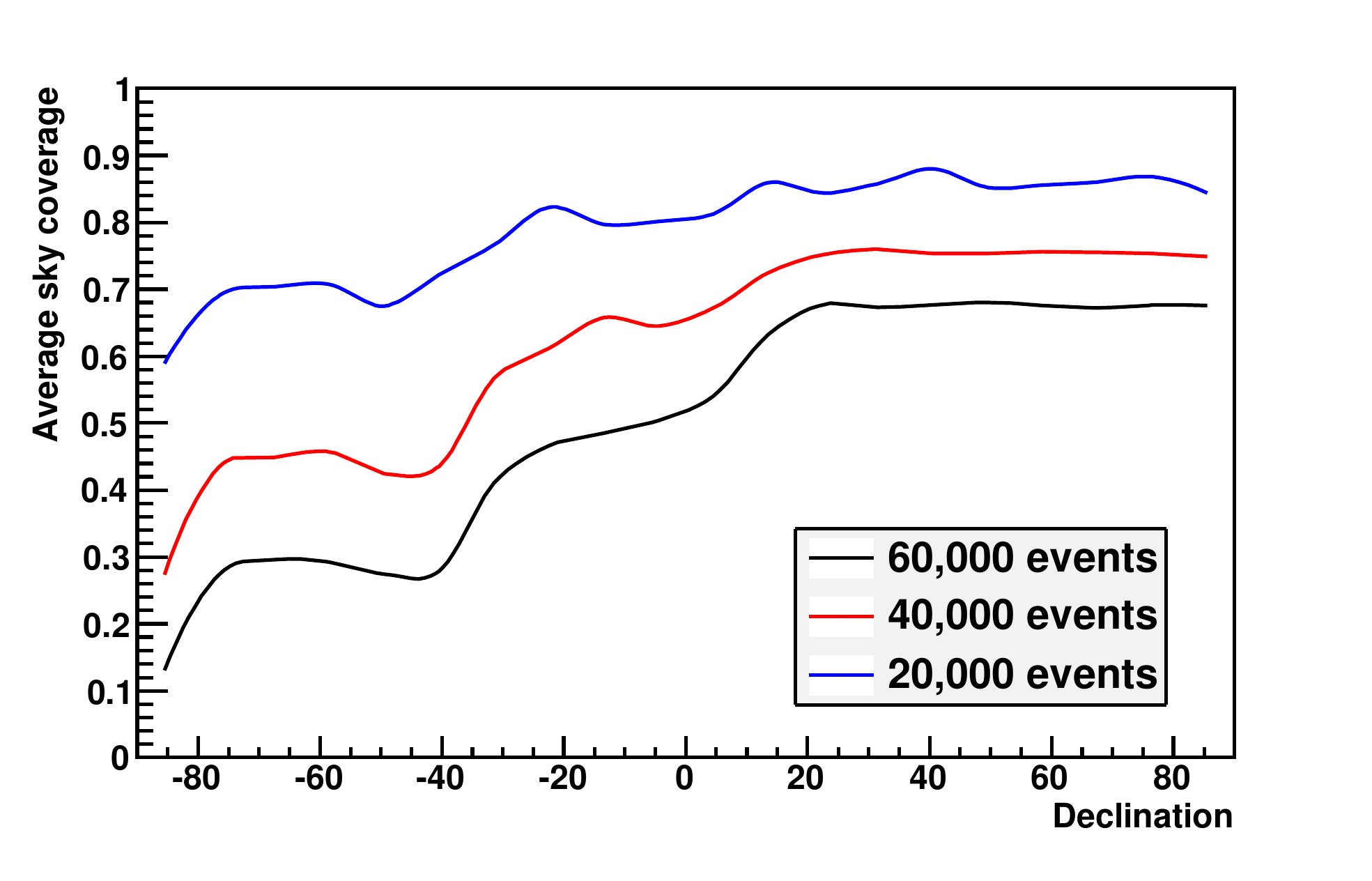}
\caption{Average scintillator sky coverage vs declination for a single detector located in Finland, for 20,000 event, 40,000 and 60,000 event signals. In total 200,000 supernovae have been simulated for the 60,000 events case and 12,000 and 15,000 for the 20,000 and 40,000 event cases, respectively.}
\label{fig:sky_coverage_vs_declination_scint}
\end{centering}
\end{figure}

Fig.~\ref{fig:scintillator_skymap} shows an example skymap for a
scintillator detector located in Finland.  Fig.~\ref{fig:sky_coverage_vs_declination_scint} shows average sky coverage vs. declination for three examples
of event statistics.

\begin{figure}[!htbp]
\begin{minipage}{3in}
\includegraphics[width=3.2in]{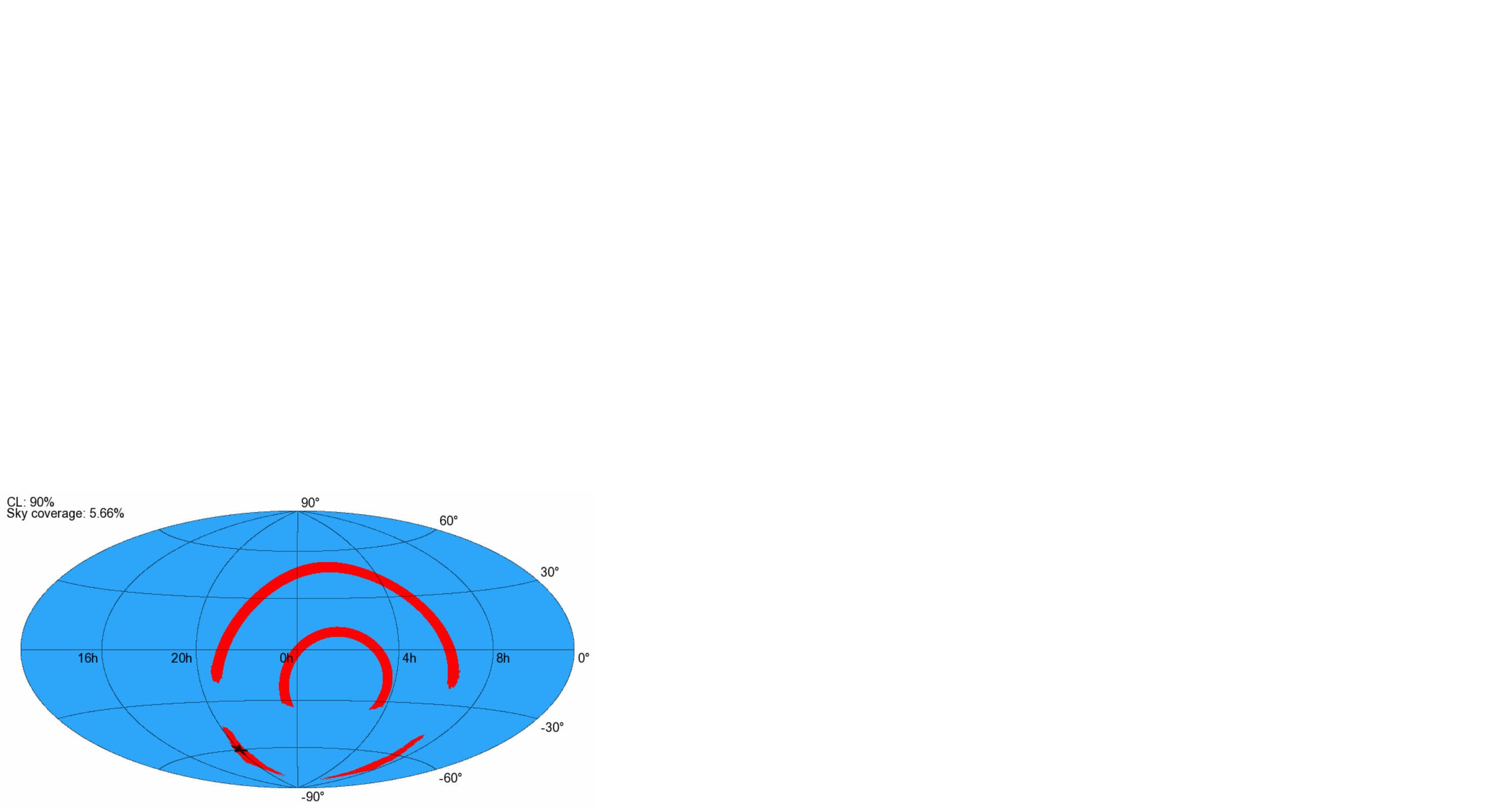}
\end{minipage}
\begin{minipage}{3in}
\includegraphics[width=3.2in]{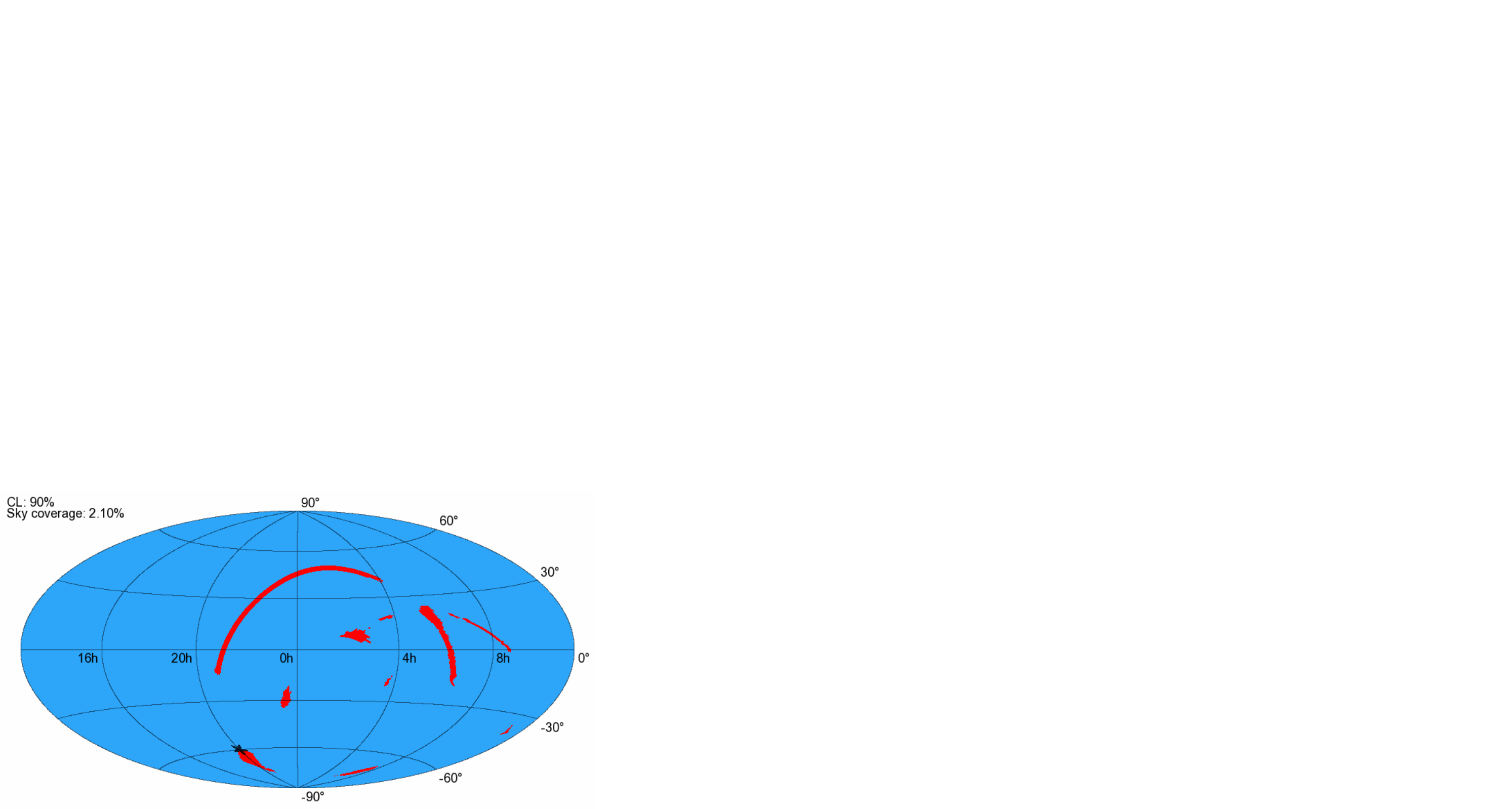}
\end{minipage}

\caption{Example scintillator skymaps, for a two (top) and three (bottom) detectors setup, each with a 60,000 events signal per detector.}
\label{fig:scintillator_skymap_multiple}
\end{figure}

\begin{figure}[!htbp]
\begin{centering}
\includegraphics[width=3.2in]{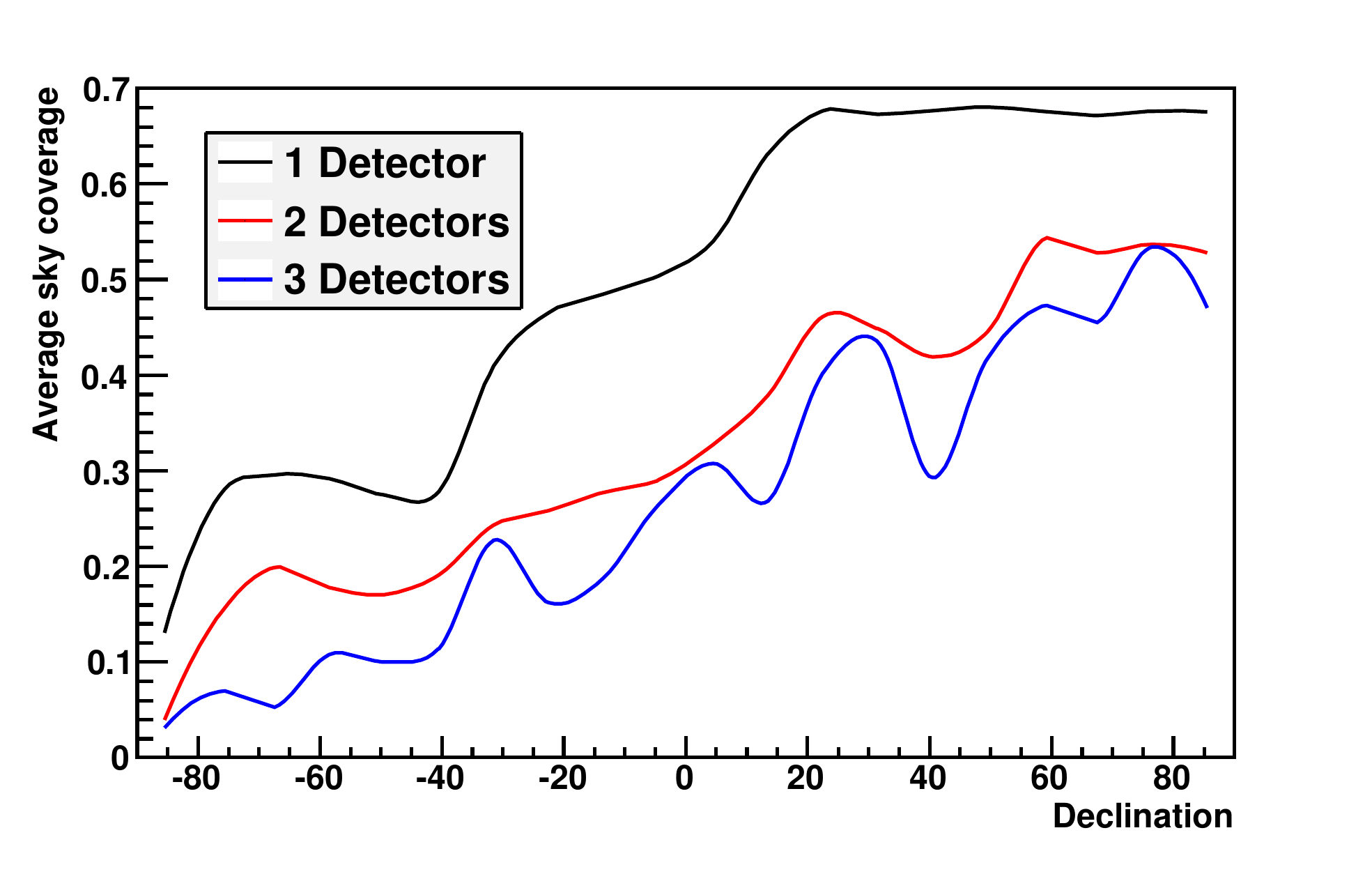}
\caption{Average scintillator sky coverage vs declination for one, two and three detectors with 60,000 events each. For one detector 200,000 supernovae have been simulated, for two detectors the number is 3,500 and for three detectors it is 1,256.}
\label{fig:sky_coverage_vs_declination_scint_multiple}
\end{centering}
\end{figure}

Next we consider the case when multiple detectors are operating: 
Fig.~\ref{fig:scintillator_skymap_multiple} shows the results of combining the
information from two and three scintillator detectors located in Finland, off the coast
of Hawaii ($19{.}72^\circ$ N, $156{.}32^\circ$ W)  
and South Dakota ($44{.}45^\circ$ N, $103{.}75^\circ$ W).  Fig.~\ref{fig:sky_coverage_vs_declination_scint_multiple} shows average sky coverage vs. declination for these
configurations.

\subsubsection{Incorporating Relative Timing Information}

We consider briefly now the possibility of incorporating relative
timing information between detectors to break degeneracies in the
allowed region(s).  A detailed study of the triangulation capabilities
for specific neutrino signal and detector models is beyond the scope of this
work.  We instead do some back-of-the-envelope estimates based on those
in reference~\cite{Beacom:1998fj}.  For a signal registered in two
detectors, the supernova direction can be constrained to a ring on the
sky at angle $\theta$ with respect to the line between the detectors,
with $\cos \theta = \Delta t/d$ and width $\delta (\cos \theta) \sim
\frac{\delta (\Delta t)}{d}$, where $d$ is the distance between the
detectors and $\delta (\Delta t)$ is the time shift uncertainty between the
pulses. We assume $\delta (\Delta t) \sim 30 {\rm~ms}/\sqrt{N_1}$, where $N_1$
is $\sim$1\% of the total signal.
A sharp feature in the signal timing could reduce $\delta(\Delta t)$.

\begin{figure}[!htbp]
\includegraphics[width=3.2in]{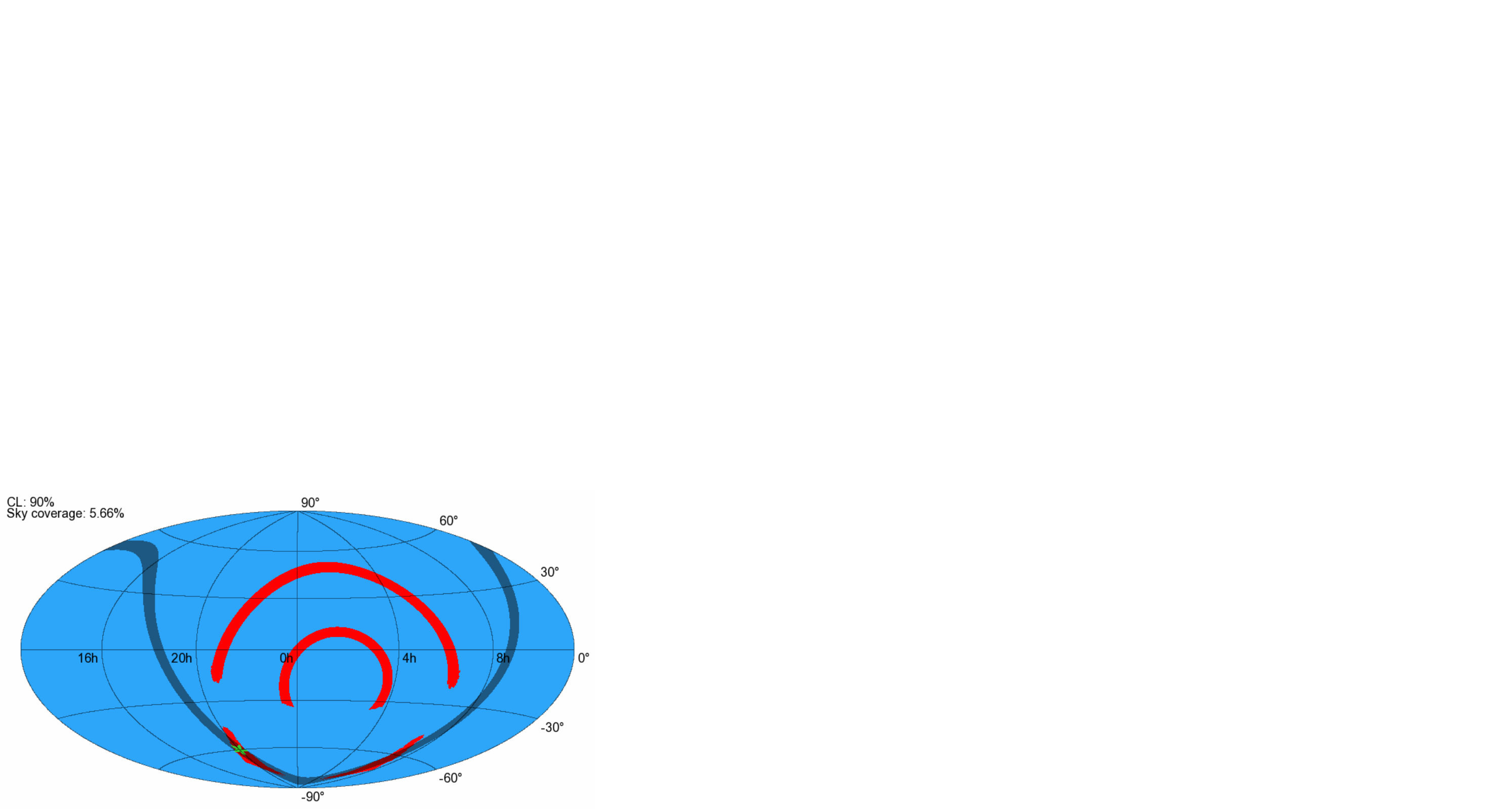}

\caption{Example two scintillator detector skymap, with estimate of allowed
  region based on relative timing information superimposed (dark band). }

\label{fig:skymap_timing}
\end{figure}

Fig.~\ref{fig:skymap_timing} shows an example for two detectors
(located in Finland and Hawaii), with the
time-triangulated allowed region superimposed: the intersection clearly narrows
down the allowed directions.

\begin{figure}[!htbp]

\includegraphics[width=3.2in]{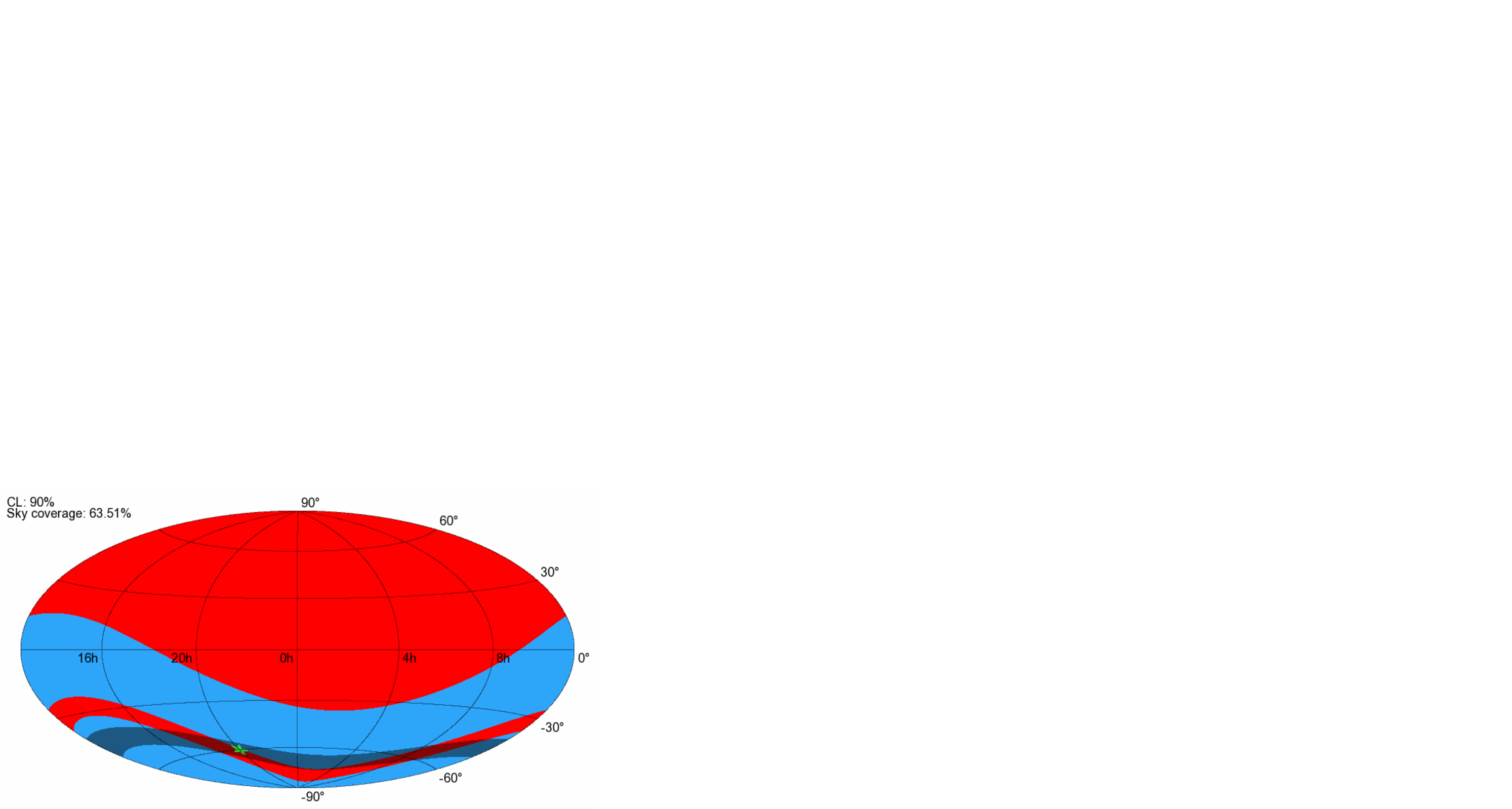}

\caption{Example single scintillator detector skymap, with estimated allowed region
determined from relative timing with the IceCube signal (dark band).}
\label{fig:skymap_timing_icecube}
\end{figure}

We can imagine also that another, non-scintillator, neutrino detector 
(or even a gravitational wave detector, \textit{e.g.}~\cite{Pagliaroli:2009qy}) could provide
relative timing information as well.  For example, IceCube 
at the South Pole could
yield few~ms timing~\cite{Halzen:2009sm}.
Fig.~\ref{fig:skymap_timing_icecube} shows an example of the intersection
of the estimated IceCube plus single scintillator detector time-triangulation 
allowed region (assuming $\delta (\Delta t) \sim 1~$ms)
with the single scintillator oscillation pattern region.

\section{Discussion}

\begin{figure}[!htbp]

\includegraphics[width=3.2in]{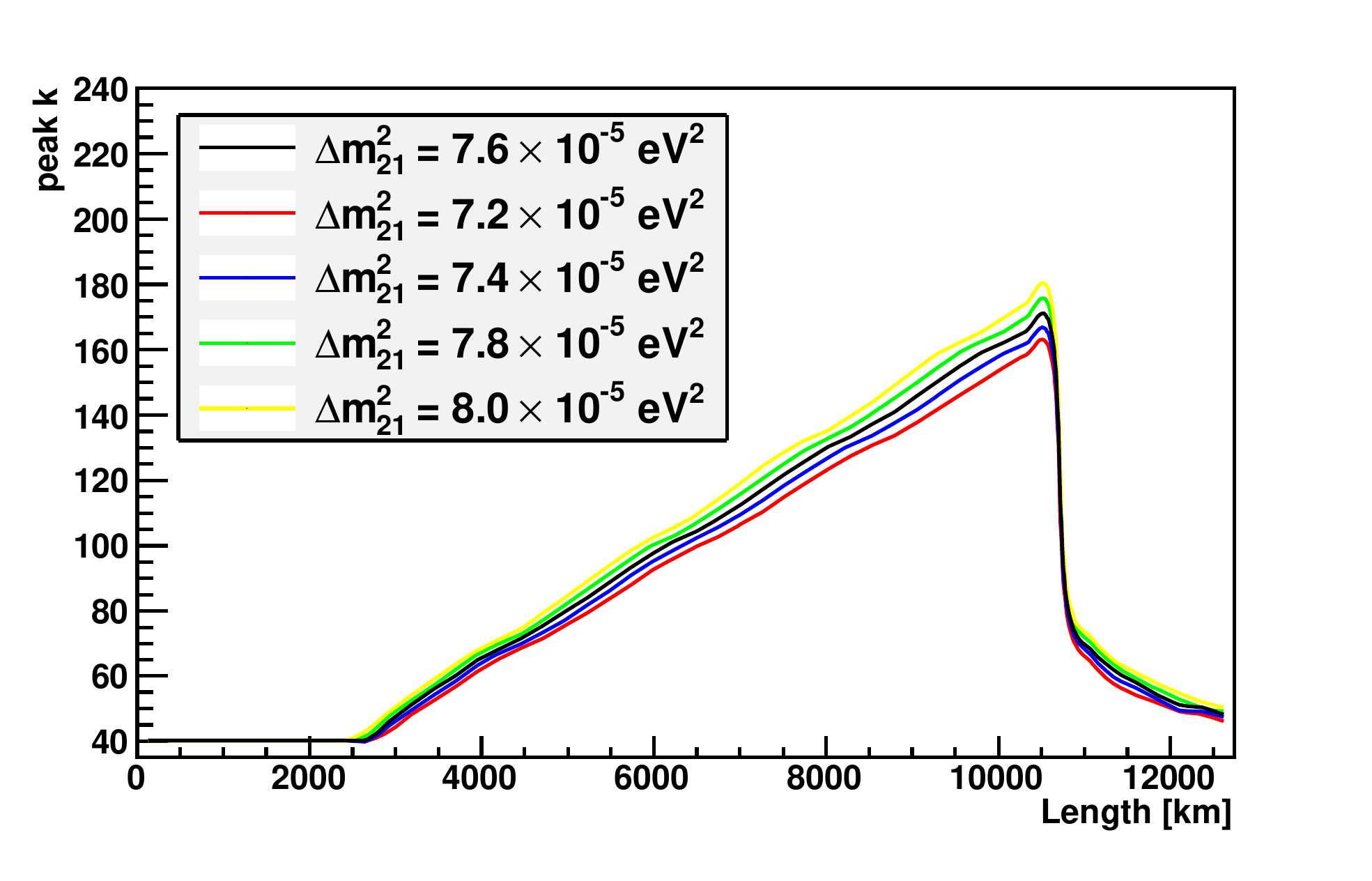}

\caption{Effect of varying  $\Delta m^2_{12}$ on the peak position 
as a function of $L$; in each case other oscillation 
parameters are held at their nominal values.}
\label{fig:parameffect1}
\end{figure}

\begin{figure}[!htbp]

\includegraphics[width=3.2in]{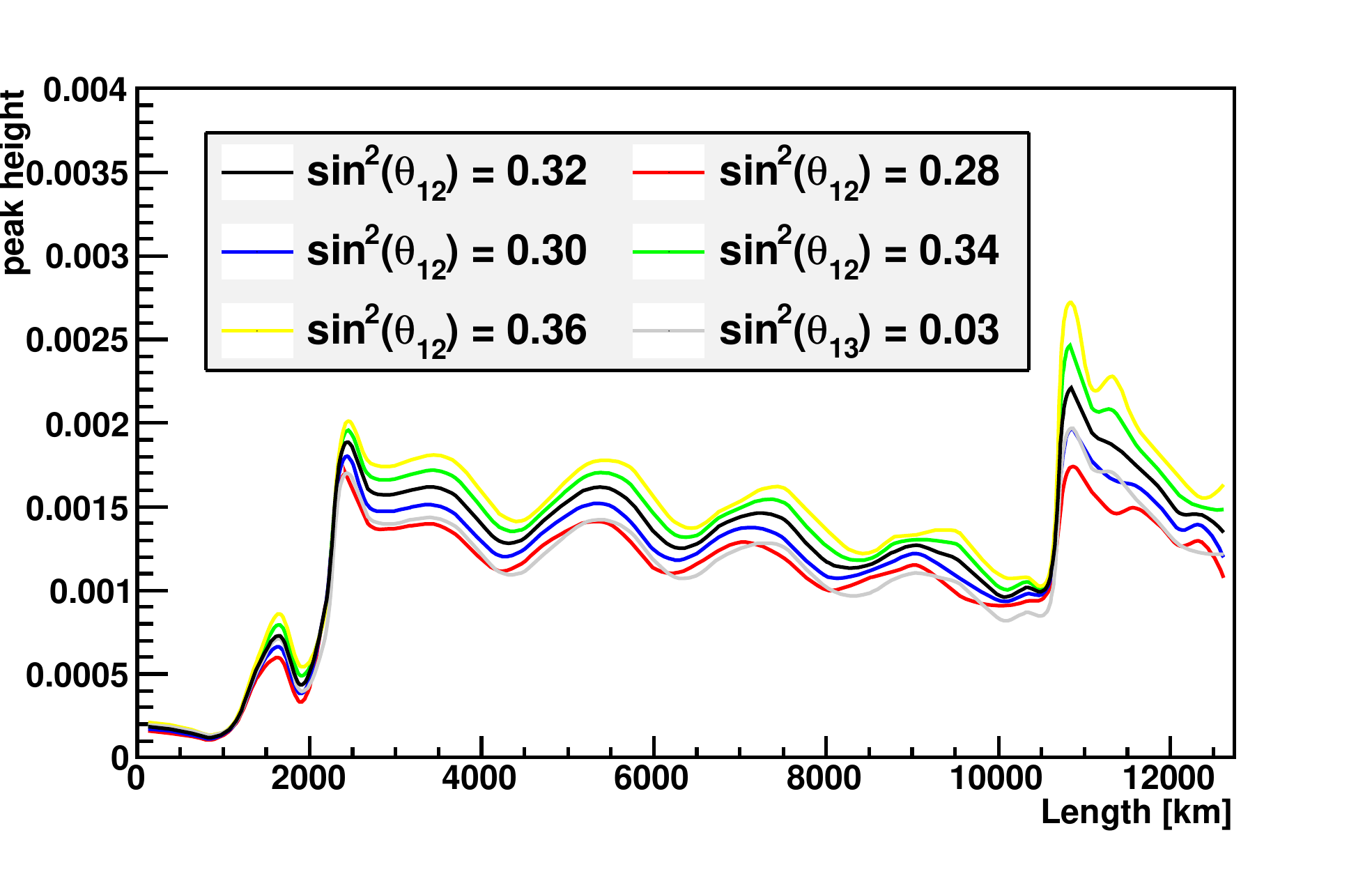}

\caption{Effect of varying $\theta_{12}$ 
and $\theta_{13}$ on the peak height 
as a function of $L$; in each case the other oscillation parameters are
held at their nominal values. }
\label{fig:parameffect2}
\end{figure}

We have assumed in these idealized
scenarios perfect knowledge of oscillation parameters.
In practice, imperfect knowledge of the oscillation parameters will
create some uncertainties.  In particular, the power spectrum
peak position is sensitive to the value of $\Delta m^2_{12}$;
the peak height is sensitive to both $\Delta m^2_{12}$ and $\theta_{12}$;
$\theta_{13}$ also has an effect on both $k_{\rm peak}$ and $h$. (The
oscillation pattern is quite insensitive to the 23 mixing parameters.)
Figs.~\ref{fig:parameffect1} and 
\ref{fig:parameffect2} show the effect on
$k_{\rm peak}$ and $h$ values of varying the oscillation parameters
within currently allowed ranges~\cite{Abe:2008ee}.

From these plots one can infer that $\lsim$1\% knowledge of the mixing
parameters is desirable.  However, one can be quite optimistic that such precision
will have been attained by the time a core collapse 
supernova happens when a large
scintillator detector is running.

Another uncertainty that will affect the quality of pointing is that
of the density of matter in the Earth.  We found only small
differences in $k_{peak}$ and $h$ from varying the mantle density by
$\pm 3\%$, or from varying the overall density by $\pm 5\%$, but
observed some changes in peak pattern for the case of neutrinos
passing through the core when varying the core density by $\pm 10\%$.

\begin{figure}[!htbp]
\begin{centering}
\includegraphics[width=3.2in]{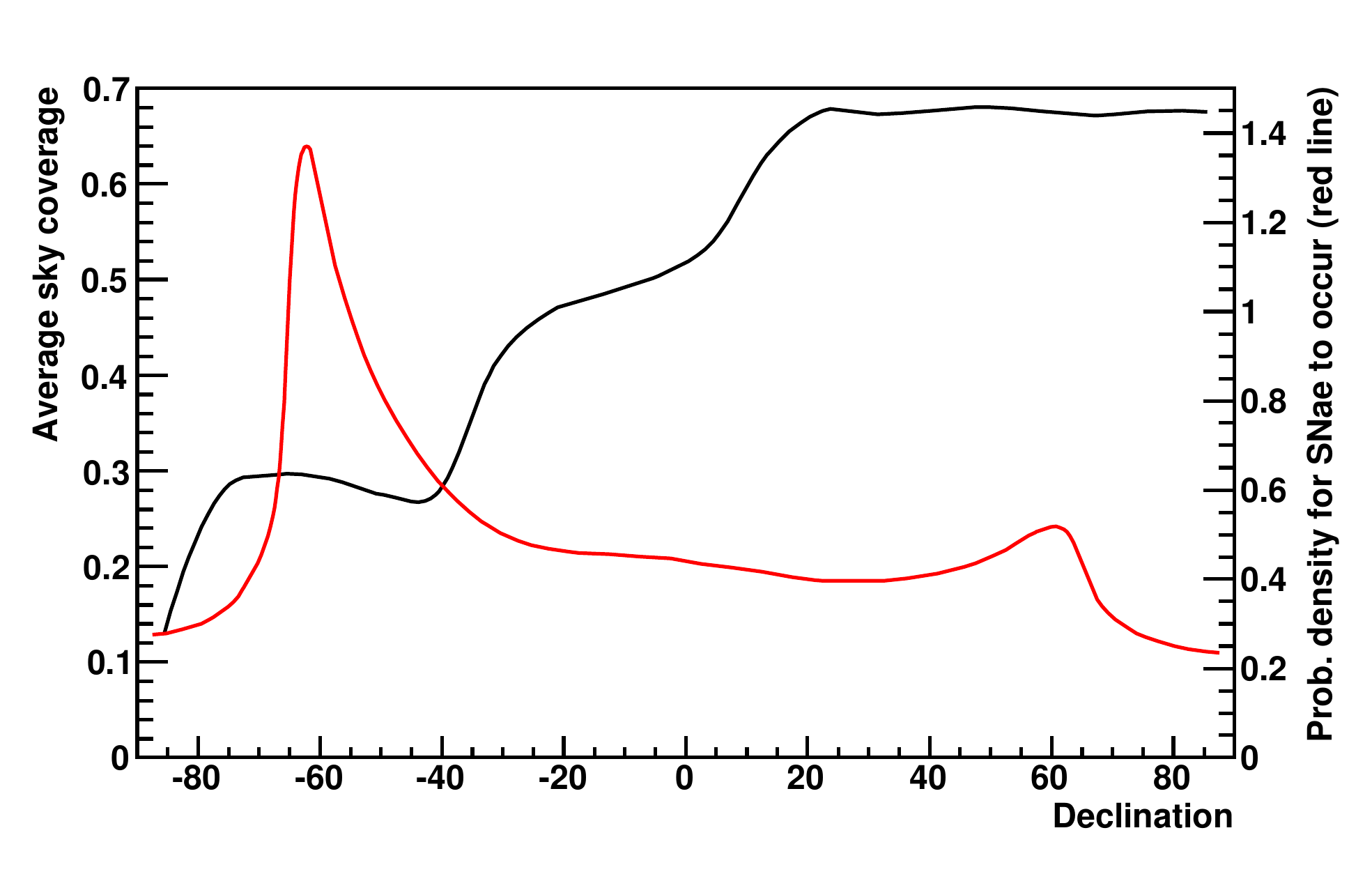}
\caption{Average scintillator sky coverage vs declination 
for a single detector in Finland, with the expected
probability for supernova occurrence superimposed 
from reference~\cite{Mirizzi:2006xx}.}
\label{fig:sky_coverage_vs_declination_with_mirizzi}
\end{centering}
\end{figure}

Many other effects may degrade the quality of direction information
that can be obtained using this technique.  There may be real spectral
features (\textit{e.g.} ``splits'') which introduce additional Fourier
components that could mask the peak, and detector imperfections may do
the same.  We acknowledge also that there may be practical
difficulties with the rapid exchange of information between
experimenters required for prompt extraction of directional
information from multiple detectors.  Nevertheless this technique
represents an interesting possibility-- even half the sky is
better than no directional information.  The oscillation pattern gives
information about direction with even a single detector, and enhances
any multiple-detector time-triangulation information.  Even if
information from only a single detector is available, or if there are
significant ambiguities, one can imagine also looking at the
intersection of the allowed region with the Galactic plane regions for
which supernovae are most likely to occur (\cite{Mirizzi:2006xx})
(perhaps using the known probability distribution as a Bayesian prior)
to improve the chances of finding the supernova: see
Fig.~\ref{fig:sky_coverage_vs_declination_with_mirizzi}.

We note that these estimates of pointing quality have been
done using a fairly simple technique based on only two parameters characterizing
the power spectra.  One can imagine employing more sophisticated
algorithms, \textit{e.g.} making use of secondary peaks or matching to
a template, and possibly incorporating knowledge of specific detector
properties or neutrino flux spectral features.  So although real
conditions may degrade quality, with this simplified study we have not
fully exploited all potentially available information.  

As a final note:
the technique could in principle work to determine directional
information for neutrino signals from other astrophysical sources,
such as black hole-neutron star mergers~\cite{Caballero:2009ww},
assuming sufficient statistics.

\section{Summary}\label{summary}

We have explored a technique by which experiments with good energy
resolution can determine information about the direction of a
supernova via measurement of the matter oscillation pattern.  This
method will only work for favorable (but currently allowed)
oscillation parameters; it requires large statistics, good energy
resolution, and well-known oscillation parameters, and it works best
for relatively long neutrino pathlengths through the Earth.  The
method is especially promising for scintillator detectors.  The
criteria will be fulfilled in optimistic but not inconceivable
scenarios.  Combining information from multiple detectors, and
possibly incorporating relative timing information, may provide
significant improvement.  The method is inferior to that using elastic
scattering in imaging Cherenkov (or argon time projection chamber)
detectors; elastic scattering remains the best bet for pointing to the
supernova.  However it is possible that a supernova will occur when no
such detector is running, in which case one should use whatever
directional information can be extracted from the observed signals.

\begin{acknowledgments}
    The research activities of KS and RW
are supported by the U.S. Department of Energy and the National Science
Foundation.  AB was supported for work at Duke University
by the Deutscher Akademischer Austausch Dienst summer internship program.

\end{acknowledgments}

\bibliography{refs}

\end{document}